\documentclass[10pt, preprint]{aastex}
\usepackage{natbib}
\slugcomment{Accepted by the \emph{Astrophysical Journal}}
\singlespace

\shortauthors{Bennett et al.}

\shorttitle{WMAP Foregrounds}

\newcommand{\iMAP}         {{\sl WMAP}}

\newcommand{\lt}           {\mbox{$<$}}
\newcommand{\gt}           {\mbox{$>$}}

\newcommand{\ddeg}         {\mbox{${\rlap.}^\circ$}}

\begin{document}

\title{First Year {\sl Wilkinson Microwave Anisotropy Probe} ({\it WMAP}) Observations: Foreground Emission}

\author{
C. L. Bennett \altaffilmark{2}, 
R. S. Hill \altaffilmark{3},
G. Hinshaw \altaffilmark{2}, 
M. R. Nolta \altaffilmark{4},
N. Odegard \altaffilmark{3},
L. Page \altaffilmark{4},
D. N. Spergel \altaffilmark{5},
J. L. Weiland \altaffilmark{3},
E. L. Wright \altaffilmark{6},
M. Halpern \altaffilmark{7},
N. Jarosik \altaffilmark{4},
A. Kogut \altaffilmark{2}, 
M. Limon \altaffilmark{2,8}, 
S. S. Meyer \altaffilmark{9},
G. S. Tucker \altaffilmark{2,8,10},
E. Wollack \altaffilmark{2}
}

\altaffiltext{1}{\iMAP\ is the result of a partnership between Princeton 
                 University and NASA's Goddard Space Flight Center. Scientific 
                 guidance is provided by the \iMAP\ Science Team.}
\altaffiltext{2}{Code 685, Goddard Space Flight Center, Greenbelt, MD 20771}
\altaffiltext{3}{Science Systems and Applications, Inc. (SSAI), 10210 Greenbelt Road, Suite 600 Lanham, Maryland 20706}
\altaffiltext{4}{Dept. of Physics, Jadwin Hall, Princeton, NJ 08544}
\altaffiltext{5}{Dept of Astrophysical Sciences, Princeton University, Princeton, NJ 08544}
\altaffiltext{6}{UCLA Astronomy, PO Box 951562, Los Angeles, CA 90095-1562}
\altaffiltext{7}{Dept. of Physics and Astronomy, University of British Columbia, Vancouver, BC  Canada V6T 1Z1}
\altaffiltext{8}{National Research Council (NRC) Fellow}
\altaffiltext{9}{Depts. of Astrophysics and Physics, EFI and CfCP, University of Chicago, Chicago, IL 60637}
\altaffiltext{10}{Dept. of Physics, Brown University, Providence, RI 02912}

\email{Charles.L.Bennett@NASA.gov}

\begin{abstract}
The \iMAP\ mission has mapped the full sky to determine 
the geometry, content, and evolution of the universe.  Full sky maps 
are made in five microwave frequency bands to separate the temperature anisotropy of the 
cosmic microwave background (CMB) from foreground emission,
including diffuse Galactic emission and Galactic and extragalactic point sources.  
We define masks that excise regions of high foreground emission, so 
CMB analyses can be carried out with minimal foreground contamination.  
We also present maps and spectra 
of the individual emission components, 
leading to an improved understanding of Galactic astrophysical processes.  
The effectiveness of template fits to remove foreground emission from the \iMAP\ data 
is also examined.  These efforts result in 
a CMB map with minimal contamination and a demonstration that the \iMAP\ CMB power 
spectrum is insensitive to residual foreground emission.

We use a Maximum Entropy Method to construct a model of the Galactic emission 
components.  The observed total Galactic emission matches the model to $\lt 1\%
$ and the individual model components are accurate to a few percent.  
We find that the Milky Way resembles other normal spiral galaxies 
between 408 MHz and 23 GHz, with a synchrotron spectral index that is 
flattest ($\beta_s \sim -2.5$) near star-forming regions, especially in the plane, 
and steepest ($\beta_s \sim -3)$ in the halo.  
This is consistent with a picture of relativistic cosmic ray electron 
generation in star-forming regions and diffusion and convection within the plane.  
The significant synchrotron index steepening out of the 
plane suggests a diffusion process in which the 
halo electrons are trapped in the Galactic potential long enough to suffer synchrotron 
and inverse Compton energy losses and hence a spectral steepening.  
The synchrotron index is steeper in the \iMAP\ bands than in lower frequency radio 
surveys, with a spectral break near 20 GHz to 
$\beta_s<-3$.  The modeled thermal dust spectral index is also steep in the \iMAP\ bands, 
with $\beta_d \approx 2.2$.  
Our model is driven to these conclusions by the low level 
of total foreground contamination at $\sim 60$ GHz. Microwave and H$\alpha$ measurements 
of the ionized gas agree well with one another at about the expected levels.
Spinning dust emission is limited to  $<\atop^{\sim}$5\% 
of the Ka-band foreground emission, assuming a thermal dust distribution with a 
cold neutral medium spectrum and a monotonically decreasing
synchrotron spectrum.

A catalog of 208 point sources is presented.  The reliability of the catalog is 98\%, i.e., 
we expect five of the 208 sources to be statistically spurious.  
The mean spectral index of the point sources  
is $\alpha\sim 0$ ($\beta\sim -2$).  Derived source counts suggest a contribution to 
the anisotropy power from unresolved sources 
of $(15.0\pm 1.4)\times 10^{-3}\;\mu{\rm K}^2$sr at Q-band and negligible levels at V-band and 
W-band.  The Sunyaev-Zeldovich effect is shown to be a negligible ``contamination'' to the
maps.

\end{abstract}



\section{INTRODUCTION}\label{intro}

The {\sl Wilkinson Microwave Anisotropy Probe} (\iMAP)$^1$ mission was designed to make precise, accurate, 
and reliable measurements of the microwave sky to allow the extraction of 
cosmological information from
the cosmic microwave background (CMB) radiation \citep{bennett/etal:2003}.  
Use of the CMB for cosmology is 
limited, however, by microwave foreground contamination from our Milky Way Galaxy and 
from extragalactic sources.  

The separation of the CMB and foreground signal components fundamentally relies on their 
differing spectral and spatial 
distributions.   To facilitate the separation of signal components, \iMAP\ was designed 
to map the full sky at five widely separated 
frequencies, from 23 GHz to 94 GHz (see Table \ref{tbl-1}).  
A similar approach was taken by the {\it COBE} 
mission \citep{bennett/etal:1992b}, although with only three frequency bands.  While {\it COBE} 
resulted in a 6144 pixel map, the \iMAP\ map has 3,145,728 sky pixels, with 
$3.995\times 10^{-6}$ sr per pixel.  The \iMAP\ mission is described by \citet{bennett/etal:2003},   
the optical design by \citet{page/etal:2003}, the feed horns by \citet{barnes/etal:2002}, 
and the radiometers and their frequency coverage by \citet{jarosik/etal:2003}. 

In this paper we follow the notation convention that flux density is 
$S\sim \nu^\alpha$ and antenna temperature is $T\sim \nu^\beta$, where the spectral 
indices are related by $\beta=\alpha-2$.  
In general, the CMB is expressed in terms of thermodynamic temperature, while Galactic and
extragalactic foregrounds are expressed in antenna temperature.  Thermodynamic temperature 
differences are given by $\Delta T = \Delta T_A [(e^x-1)^2/x^2e^x]$, where $x=h\nu/kT_0$, 
$h$ is the Planck constant, $\nu$ is the frequency, $k$ is the Boltzmann constant, and 
$T_0=2.725$ K is the CMB temperature \citep{mather/etal:1999}.  
A band-by-band tabulation of the conversion
factor is given in Table \ref{tbl-1}.

\begin{deluxetable}{lccccc}
\tablecaption{Conversion Factors and Selected Values by \iMAP\ Band  \label{tbl-1}}
\tablewidth{0pt}
\tablehead{
\colhead{} & \colhead{K-Band\tablenotemark{a}}   & \colhead{Ka-Band\tablenotemark{a}}   &
\colhead{Q-Band\tablenotemark{a}} &
\colhead{V-Band\tablenotemark{a}}  & \colhead{W-Band\tablenotemark{a}}
}
\startdata
Wavelength (mm)\tablenotemark{a}        &   13   &   9.1   &  7.3     &   4.9  &  3.2 \\
Frequency (GHz)\tablenotemark{a}        &   23   &    33   &  41      &   61   &  94  \\
$\Delta T / \Delta T_A$                 & 1.014  &  1.029  &  1.044   & 1.100  & 1.251 \\
T$_{\rm A}$/I(R) ($\mu$K R$^{-1}$)       & 11.4   &  5.23   &  3.28    & 1.40   &  0.56 \\
$T_A/EM$ ($\mu$K per cm$^{-6}$ pc)      &  4.95  &   2.29  &  1.44    &  0.614 & 0.242 \\
\enddata

\tablenotetext{a}{Typical values for a radiometer are given.   
See text, \citet{page/etal:2003}, and \citet{jarosik/etal:2003} for exact values, 
which vary by radiometer.}

\end{deluxetable}

Many sky regions are so strongly contaminated by foregrounds that a clean 
separation of the CMB 
is impossible, so in \S\ref{mask} we define a pixel mask to be used for CMB analyses.  
Microwave and other measurements show that, at high Galactic
latitudes ($\left| b\right| >15^\circ$), CMB anisotropy dominates the Galactic 
signals in the frequency range $\thicksim 30-150$ GHz 
\citep{tegmark/etal:2000, tegmark/efstathiou:1996}. This fact makes the use 
of the masks discussed in \S\ref{mask} attractive.  

In \S\ref{mechanisms} we address physical emission mechanisms that contribute to 
Galactic emission: synchrotron, free-free, thermal dust, and nonthermal (spinning or 
thermally fluctuating magnetic) dust emission.
In \S\ref{ilc} we demonstrate a method that combines the \iMAP\ maps in such a way as 
to isolate the CMB.  In \S\ref{fitting} we separate the signals by emission 
type to better understand the foregrounds.  Through a better understanding
of the foregrounds we are able to demonstrate the degree of residual 
contamination for CMB analyses after masks are applied.  We also  
improve our understanding of the astrophysics of the foreground emission.
In \S\ref{templatefits} we prepare a set of maps for cosmological analysis:  
we apply the masks from \S\ref{mask} to remove the brightest regions of emission, 
and then remove residual foreground emission using the templates introduced in 
\S\ref{mechanisms}.  In \S\ref{pointsources} we estimate the level of 
extragalactic point source 
contamination and provide a list of the strongest sources.  In \S\ref{sz} we assess the 
degree to which the WMAP maps are affected by the Sunyaev-Zeldovich effect.  We 
conclude in \S\ref{conclusions} by summarizing the major results of the paper.

\setcounter{footnote}{0}
\section{FOREGROUND MASKS \label{mask}}

The purpose of a foreground mask is to exclude map pixels that contain ``too much'' 
foreground signal from use in CMB analyses.  Some pixels often need to be masked 
even after a foreground reduction technique has been applied to a map.  
What constitutes ``too much'' foreground signal depends on the particular 
analysis. It is therefore useful to have masks available with different flux 
cut-off levels.   For example, these masks can be used to demonstrate (in)sensitivity
of scientific results to the level of foreground sky cut.  

The mask cut-off level is always a compromise between eliminating foregrounds 
and preserving maximal usable sky area in analyses.  The majority of 
excluded sky area in any mask is the Galactic plane region.
The mask has the same pixel format as the map, where each mask pixel 
indicates whether or not the corresponding map pixel is to be used in analyses.  

\begin{figure}
\figurenum{1}
\epsscale{1.}
\plotone{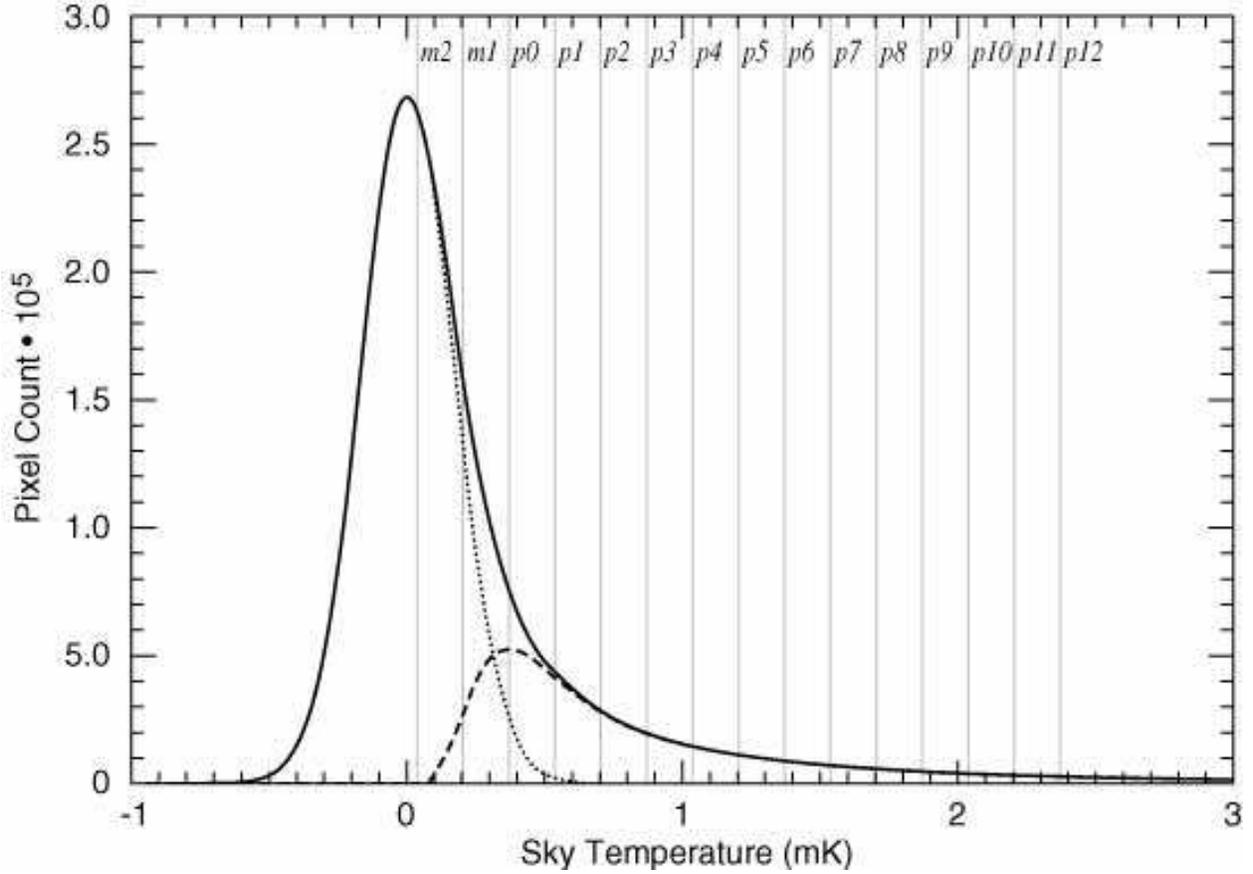}
\caption{The solid curve is a histogram of pixel values in the K-band map.  The dotted line is the 
symmetric reversal of the negative sky temperature portion of the solid curve about the peak value. The dashed curve 
is their difference or remainder, 
which is attributed to excess foreground emission.  The peak of the remainder curve 
defines the K-band sky temperature cut-off for the mask that we call ``Kp0''.  A series of masks with 
varying degrees of severity are likewise defined with ``m'' for ``minus'' and ``p'' for ``plus'', as shown. 
Temperature steps are defined by the rms half-width of the sky histogram for values less than the 
mode. \label{hist}} 
\end{figure}

A successful mask should
exclude those broad sky areas that are dominated by foregrounds, while
leaving unexpunged the sky areas dominated by the CMB.  Further, a useful 
recipe should be based on an  
unbiased process.   
\iMAP\ has adopted a small set of standard masks of varying
severity but with a common recipe.  All masks are based on the K-band map
since the foreground contamination is most severe in this lowest frequency 
band.  The histogram distribution of all the pixel
brightnesses in this map is strongly asymmetric in favor of positive pixel
values because of the foreground contamination, as seen in
Figure \ref{hist}.

We reflect the (relatively uncontaminated) left side of the observed
histogram about the mode.  This symmetrized distribution is subtracted 
from the observed distribution,
leaving a remainder, which is interpreted as contamination.
The remainder histogram defines cutoff brightnesses.

The cutoff is applied to a low-resolution version of the K-band map in
order to avoid excluding an archipelago of isolated pixels.  \iMAP\ images are
generally pixelized with a HEALPix\footnote{see http://www.eso.org/science/healpix/}  
$N_{side}$ parameter of 512,
whereas the masks are generated at $N_{side}=64$.  However, this
procedure gives the mask a peculiar boundary inherited from the
diamond-shaped pixels.  Accordingly, the boundary is 
softened by filtering the mask, which consists of ones
and zeros, with a Gaussian of $2^\circ$ full-width at half-maximum (FWHM), 
and taking the 0.5
contour of the result as the new mask boundary.

\begin{figure}
\figurenum{2}
\epsscale{1.}
\plotone{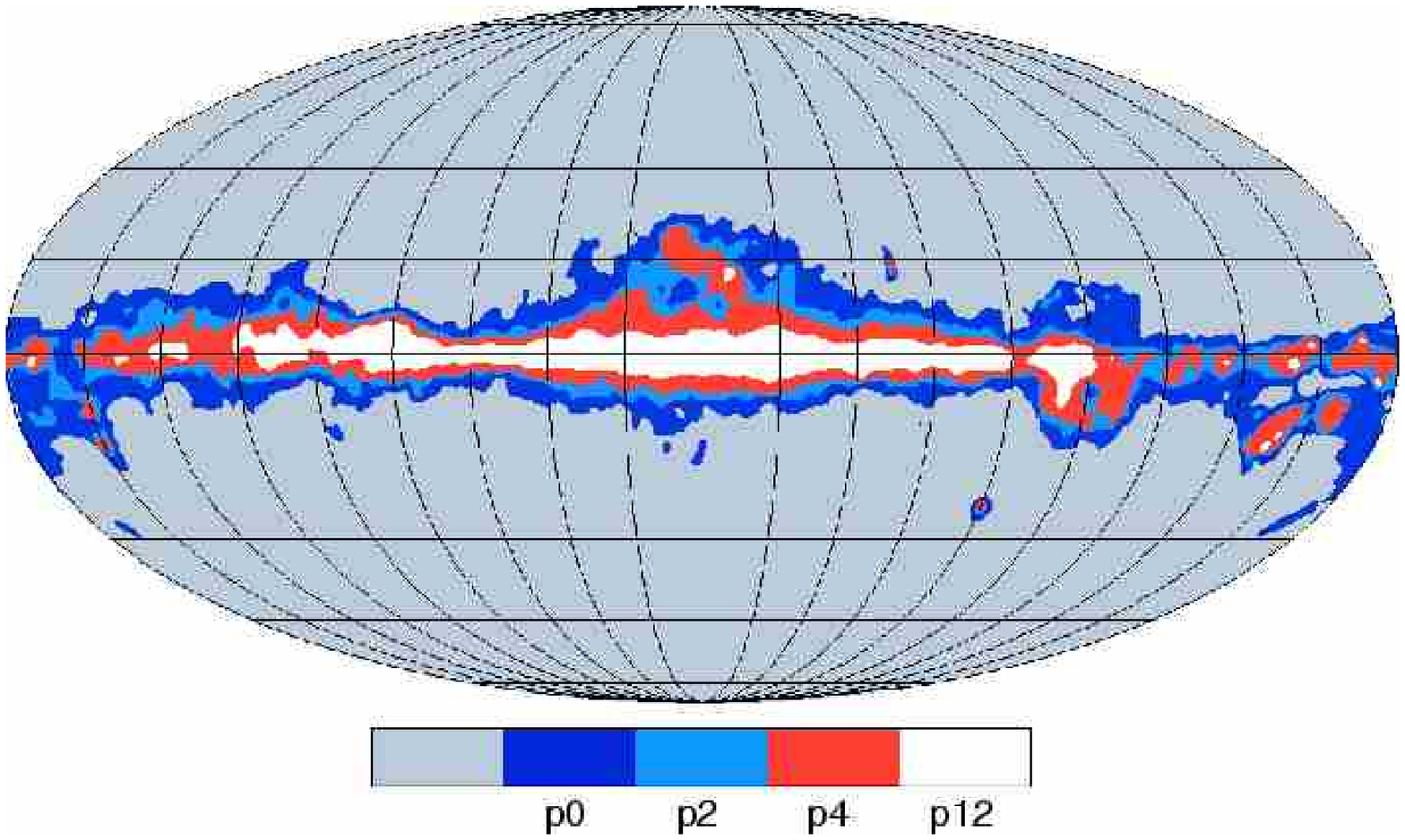}
\caption{A full sky map of selected \iMAP\ standard masks.  The masking is based on the K-band signal
levels, as discussed in \S\ref{mask}.  The series of masks allows for a choice of cut severity. 
(Each successive mask includes the previous regions; e.g., the p4 mask includes all of the pixels
in the p12 mask.)
\label{maskmap}} 
\end{figure}

As explained above, a range of cutoff levels is
desirable.  A convenient step size in cut-off temperature is derived from the
sky histogram, shown in Figure \ref{hist}, by using the root-mean-square
(rms) half-width of this
histogram for values less than the mode.  A series of masks are made using 
multiples of these step sizes.  A cut at the peak value of the remainder 
histogram is referred to as ``p0''.  The sky cut used in any analysis is 
described by a symbolic name.  For example, the 
mask defined by the peak of the remainder histogram using the K-band map is ``Kp0''.
A mask with a temperature cut-off two steps smaller in temperature (more severe) 
is called ``Km2'', where ``m'' means ``minus'' and (for a less severe cut) 
``p'' means ``plus''.   
Figure \ref{maskmap} shows selected masks based on the definitions 
in Figure \ref{hist}.  A separate set of masks based on bands other than K-band is not 
necessary since they correspond closely with specific K-band defined masks.  

In addition to diffuse Galactic emission, point sources also contaminate the \iMAP\ maps.  
We have constructed a point source source mask that includes: (1) all of the sources from 
\citet{stickel/meisenheimer/kuehr:1994} (which is the updated \citet{kuehr/etal:1981} 
5 GHz 1 Jy compilation with attached 
optical source identifications); (2)  sources with 22 GHz fluxes $\ge 0.5$ Jy from the 
\citet{hirabayashi/etal:2000} comprehensive compilation of flat spectrum AGN; (3)  Additional flat 
spectrum objects from \citet{terasranta/etal:2001}; and (4) sources from the X-ray/radio 
blazar survey of \citet{perlman/etal:1998} and  \citet{landt/etal:2001}. 
The mask excludes a radius of $0.6^\circ$ about each of almost 700 source positions. 
Table \ref{tbl-2} gives the percentages of the sky that are cut by each of the analysis masks
with and without masking point sources.

\begin{deluxetable}{ccc}
\tablecaption{Percent Area Masked by Sky Cut Level.\label{tbl-2}}
\tablewidth{0pt}
\tablehead{
\colhead{Specification} & \colhead{\% of pixels cut} & \colhead{\% of pixels cut}\\
                        &                            & \colhead{including point sources}
}
\startdata
Kp0  &  21.4 & 23.2 \\
Kp2  &  13.1 & 15.0 \\
Kp4  &   9.3 & 11.2 \\
Kp12 &   3.8 &  5.8 \\
\enddata
\end{deluxetable}
  
\section{GALACTIC MICROWAVE EMISSION \label{mechanisms}}

We have two objectives.  One is to fit and remove the foregrounds to derive CMB maps 
with well-specified levels of contamination.  The second is to
advance our understanding of the astrophysics of the foreground emission.  
In support of these two objectives, this section summarizes the astrophysical microwave 
emission mechanisms and relevant existing measurements. 

\subsection {Free-free emission \label{ff}}

Free-free emission arises from electron-ion scattering.
Free-free thermal emission has a $T_A\sim \nu^\beta$ high-frequency ($\nu > 10$ GHz) spectrum, 
with $\beta=-2.15$,
and a low-frequency rise of $T_A\sim \nu^2$ due to optically thick self-absorption.  

Radio astronomy provides no free-free emission map of the sky because free-free 
doesn't dominate the sky at any radio frequency. High-resolution, large-scale, maps of H$\alpha $
(hydrogen n=3$\rightarrow$2) emission 
\citep{dennison/simonetti/topasna:1998, haffner/etal:2002, reynolds/haffner/madsen:2002, 
gaustad/etal:2001} 
can serve as an approximate template for 
the free-free emission, except in regions of high interstellar dust optical depth, $\tau>1$, 
at the H$\alpha$ wavelength (about 16\% of the sky).  
For $\tau \le 1$, the H$\alpha$ maps can be approximately corrected for the 
effects of extinction.

For $T_e \lt 550,000$ K, the free-free volume emissivity is given by

\begin{equation}
\epsilon_\omega d\omega = {\frac{8n_en_iZ^2e^6}{3\pi \sqrt{2\pi} m^2 c^3}} \sqrt{{\frac{m}{kT_e}}}\;
\ln \left[ \left( {\frac{2 k T_e}{\psi m}} \right) ^{3/2} {\frac{2m}{\psi Ze^2 \omega}} \right] d\omega
\end{equation}

\noindent \citep{oster:1961}, where $T_e$ is the electron temperature, 
$Ze$ is the net charge of the ion, $\psi=1.78$ is from 
Euler's constant, $m$ and $e$ are the mass and charge of the electron, 
$\omega$ is the angular frequency of
emission, and $n_e$ and $n_i$ are the electron and ion volume densities.  
We take $n_e=n_i$ and, for 
$T_e\approx 8000$ K (see, e.g., \citet{otte/gallagher/reynolds:2002}), 
it is safe to approximate $Z=1$.  We then use the 
definition of the emission measure, $EM= \int n_e^2 dl$, to get an expression for the \iMAP\ 
free-free antenna temperature,

\begin{equation}
T_A^{WMAP}(\mu {\rm K}) = 1.44 EM_{{\rm cm}^{-6}{\rm pc}} \times 
\frac{[1 + 0.22\; \ln (T_e/8000 \;{\rm K}) - 0.14\; \ln (\nu/41 \;{\rm GHz})]}
{(\nu/41 \;{\rm GHz})^2(T_e/8000 \;{\rm K})^{1/2}}
.
\end{equation}

\noindent Values of $T_A/EM$ for the 5 \iMAP\ bands are given in Table \ref{tbl-1}, 
assuming $T_e=8000$ K.

The free-free spectral index for \iMAP\ frequencies,

\begin{equation}
\beta^{WMAP}_{ff} = -2 - \frac{1}{10.48 + 1.5\; \ln (T_e/8000 {\rm K}) - \ln\, \nu_{\rm GHz}}
\end{equation}

\noindent does not extend to arbitrarily low frequency.  In general,
$T_A=T_e\;(1-e^{-\tau_\nu})$.  At the \iMAP\ frequencies $T_A<<T_e$, i.e. 
$\tau_\nu^{WMAP} << 1$, so $T_A^{WMAP} \approx \tau_\nu T_e$.  Through this last relationship, 
we can use the \iMAP\ measurement of $T_A^{WMAP}$ to infer $\tau_\nu$ and thus 
extend a predicted 
antenna temperature to low frequencies where $\tau_\nu \ge 1$.  This can be done, for example, 
when comparing the \iMAP\ free-free measurements in the Galactic plane to the
\citet{haslam/etal:1981} 408 MHz observations.
The procedure is only approximate, however, since $T_e$ and the beam filling 
factors for discrete sources are not accurately known.

The intensity of H$\alpha$ emission is given by 

\begin{equation}
I({\rm R}) = 0.44\; \xi \; EM_{\rm cm ^{-6} pc}\; (T_e/8000 {\rm K})^{-0.5} 
\left[ 1 - 0.34\; \ln (T_e/8000 {\rm K}) \right],
\end{equation}

\noindent where the H$\alpha$ intensity is in units of Rayleighs 
(1 R $ = 2.42 \times 10^{-7}$ ergs cm$^{-2}$s$^{-1}$ sr$^{-1}$ at 
the H$\alpha$ wavelength of 0.6563 $\mu$m), the helium contribution is assumed to be small,
and $\xi$ is an extinction-correction factor.  If, for example, the emitting gas 
is co-extensive with  
dust, then $\xi =[1-{\rm exp}(-\tau)]/\tau$.  H$\alpha$ is in R-band where 
the extinction is 0.75 times visible, $A_R/A_V=0.75$.  Thus, $A_R=2.35 E_{B-V}$, so $\tau=2.2
E_{B-V}$. 
 
\begin{figure}
\figurenum{3}
\epsscale{0.5}
\plotone{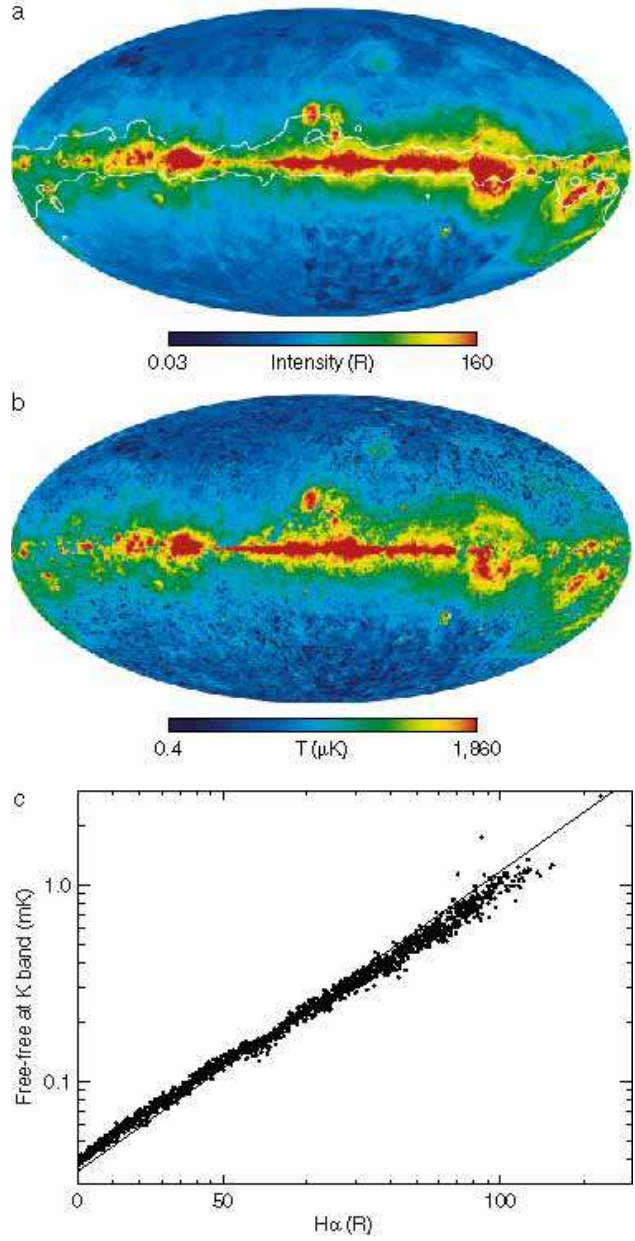}
\caption{
\footnotesize
(a) A H$\alpha$ map of the sky, corrected for extinction.  The correction is
only an approximation and is especially unreliable for regions where $\tau>1$.  These high
opacity regions are roughly demarcated by the contour lines, which have been smoothed for 
clarity. \label{Halpha}
(b) A full sky map of free-free emission based on the five bands of \iMAP\ 
measurements.  The map is the result of the MEM modeling process, as described in
the text.\label{MEMff}
(c)  Pixel values in the MEM-derived free-free map are compared against their values 
in the H$\alpha$ map, with each point representing an average in H$\alpha$ intensity bins of 
$\Delta I = 8$ R.  
The lack of scatter in the fainter bins reflects fidelity to the prior assumption of 
11.4 $\mu$K/R at K-band.  The data plotted only includes pixels where the H$\alpha$ optical depth
is estimated to be $<0.5$ and the K-band antenna temperature is $>0.2$ mK.  The data are
consistent with the $\mu$K/R values given in Table \ref{tbl-1}  within $\sim\pm 12$\%
uncertainty, depending on the fitting method used.
\label{TTMEMff}
\label{ion}} 
\end{figure}

\citet{finkbeiner:2003} assembled a full sky H$\alpha$ map using data from several 
surveys:
the Wisconsin H-Alpha Mapper (WHAM), the Virginia Tech Spectral-Line Survey (VTSS), and the 
Southern H-Alpha Sky Survey Atlas (SHASSA).
This map, 
with the \citet{schlegel/finkbeiner/davis:1998} (hereafter referred to as SFD) extinction 
map of the sky, are used together to compute a best guess map of free-free emission in regions 
where $\tau \lt 1$, under the assumption of co-extensive dust and ionized gas.  The 
resulting template is shown in Figure \ref{ion}(a).  

We note that this template only approximates the
free-free emission.  There are several sources of uncertainty and error.  For example:
(a) H$\alpha$ light scatters from dust grains in a manner unlike free-free;
(b) There is uncertainty in the calibration of the H$\alpha$ measurements, both within and 
between the observational groups;
(c) There is some error in separating the geocoronal emission from the H$\alpha$ line emission;
(d) Any inaccuracy in the assumed Balmer atomic rates will also introduce error;
(e) The H$\alpha$ intensity depends on $T_e$, and it is not known 
precisely how $T_e$ changes across the Galaxy;
(f) There is uncertainty in ionization state of helium;
(g) Any gas that has a sufficient bulk velocity could emit outside the bandwidth of the 
H$\alpha$ observations; and
(h) The extinction correction assumes that the H$\alpha$ emission 
is co-extensive with the dust along the line of sight. 

It would be difficult to assess and propagate all of the above sources of error.
A comparison with a \iMAP\ free-free map will be discussed in \S{\ref{mem}}.

\subsection{Synchrotron Emission \label{sync}}

Synchrotron emission arises from the acceleration of cosmic ray electrons in  
magnetic fields.  The acceleration occurs in Type Ib and Type II supernova remnants.
The same $M\gt 8{\rm M_\odot}$ stars that form these remnants also emit UV radiation 
that heats dust grains and ionizes hydrogen in the interstellar medium.  Thus synchrotron 
emission is associated with the star-formation cycle.

The energy spectrum of cosmic ray electrons is expressed as a relativistic 
electron number density 
distribution $N(E)\sim E^{-\gamma}$.  Since $N(E)$ varies across the galaxy, 
as does the magnetic field, ${\bf B}$, 
the resulting synchrotron emission can be characterized by a wide range of spectral 
behaviors, and hence {\it the observed morphology of synchrotron sky maps 
will change substantially with frequency}.
The synchrotron flux density spectral index, $\alpha$, is related to the electron power law 
index, $\gamma$, by $\alpha = -(\gamma-1)/2$ at frequencies greater than a few GHz, where 
self-absorption is negligible.

The synchrotron emission spectrum is further affected by cosmic ray propagation, energy loss, and
degree of confinement.  Cosmic ray electrons can propagate via diffusion and/or convection.
Diffusion involves the random scattering of cosmic ray electrons from varying magnetic fields,
while convection implies the systematic outward movement of the sources of scattering.

Cosmic rays lose energy via synchrotron loss, inverse Compton scattering, adiabatic loss,
and free-free loss.  Cosmic ray electrons may be largely confined to their host galaxies, or may 
freely leave their halos.  Highly confined cosmic ray electrons lose their energy before they 
escape from the halo of the galaxy, while poorly confined electrons escape the galaxy before 
they lose a substantial portion of their energy.
A steep spectral index (e.g., $\alpha \lt -0.9$, $\beta \lt -2.9$) 
indicates a high rate of energy loss and a low escape rate.  
A flatter spectral index (e.g., $\alpha \gt -0.7$, $\beta \gt -2.7$) 
indicates that electrons can escape their host galaxy before losing a significant fraction 
of their energy.  This can imply strong convection.

Synchrotron emission can be categorized as arising from two types of sources: electrons trapped 
in the magnetic fields of discrete supernova remnants and diffuse emission from cosmic ray 
electrons spread throughout the Galaxy.   
In a discrete supernova the magnetic field (typically $\sim 75\; \mu$G) is enhanced because it 
is frozen into the shocked compressed gas.  The diffuse magnetic field in the Galaxy is typically 
only $1-5\; \mu$G.

Discrete supernova remnants typically have a spectral index of $\alpha\sim -0.5$ (i.e.  
$\gamma\sim 2$) in the few GHz radio range \citep{green:1988} (see also 
http://www.mrao.cam.ac.uk/surveys/snrs/snrs.info.html)
and contribute only $\sim 10 \%
$ of the total synchrotron emission of the Galaxy at 1.5 GHz \citep{lisenfeld/volk:2000,
biermann:1976, ulvestad:1982}, 
despite the enhanced magnetic field strength.  More than 90\%
of the observed synchrotron emission arises from a diffuse component with a 
direction-dependent spectral
index that generally lies in the range $-0.5 \gt \alpha \gt -1.1$ ($-2.5 \gt \beta \gt -3.1$ and 
$2.0 \lt \gamma \lt 3.2$).  The cosmic ray electrons in the Galaxy substantially
outlive their original supernova remnants and slowly lose energy while traveling large 
distances across the Galaxy.
 
The nonthermal spectral index is seen to steepen off the disk in our Galaxy and 
the same effect is seen in external galaxies.  The observed steepening implies that energy 
loss is important \citep{lisenfeld/volk:2000}.    

The low-frequency cut-off of synchrotron emission arises both from the 
single-electron synchrotron spectrum ($F(x)=x \int_x^\infty{K_{5/3}(\xi)d\xi}$, where $K_{5/3}$ 
is the modified Bessel function) low-frequency cut-off, and from self-absorption and free-free 
absorption, both of which become increasingly efficient at lower frequencies. The rise in the 
low-frequency synchrotron spectrum in the optically thick regime goes as $T_A\sim \nu^{2.5}$.   
Synchrotron emission is also cut-off by the Razin-Tsytovich effect, which describes a 
strong suppression of the synchrotron 
radiation when the phase velocity of the light in the plasma is greater 
than $c$.  The synchrotron cut-off occurs at frequencies below where the kinetic temperature 
is equal to the brightness temperature.  This often occurs at a frequencies of a few MHz, but
can also play a role in GHz-peaked sources.  

Synchrotron emission can be highly polarized, with the degree of linear polarization as high 
as $(3\gamma+3)/(3\gamma+7)$, i.e. $\sim 75\%
$ polarized.  This is almost never observed,
however, due to Faraday rotation ($\propto \lambda^2$) and non-uniform magnetic field directions 
along a line of sight, which generally reduce the degree of observed polarization to $\lt 20\%
$.

\begin{figure}
\figurenum{4}
\epsscale{0.5}
\plotone{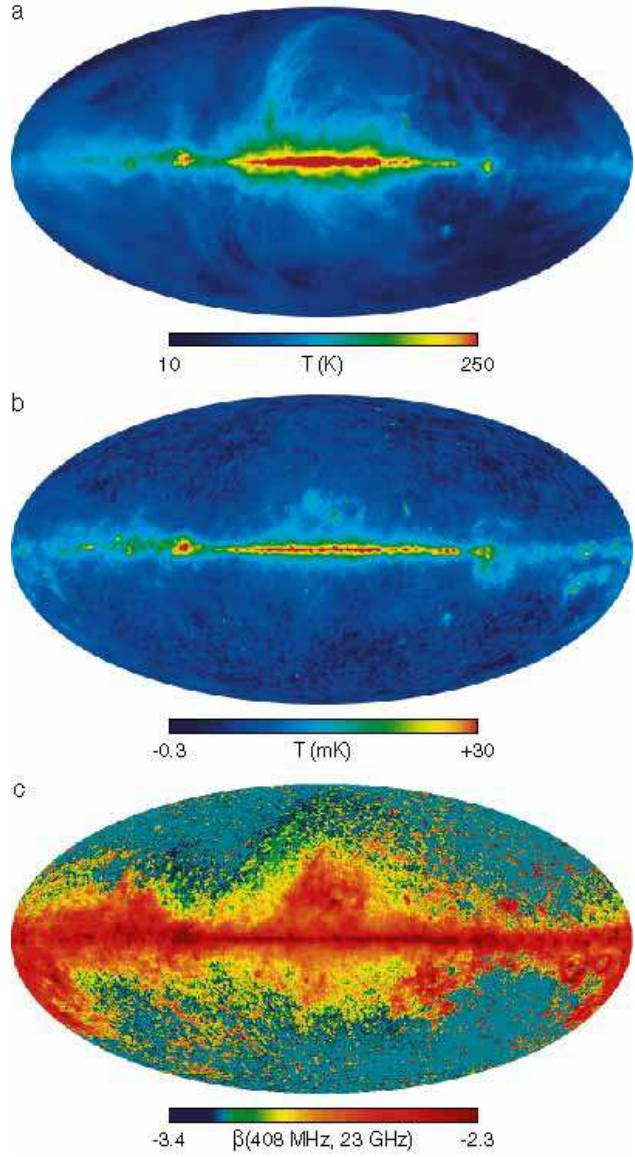}
\caption{\footnotesize
(a) The Haslam 408 MHz sky map is largely dominated by synchrotron emission.
(b) The 23 GHz K-band \iMAP\ map, also dominated by synchrotron emission, is more 
concentrated towards the plane than the 408 MHz Haslam map because flatter spectral
index regions increasingly dominate at the higher observing frequencies. The steep 
spectral index North Galactic Spur, for example, is much less apparent at \iMAP\
frequencies.  The variable synchrotron spectral index across the sky renders the Haslam 
408 MHz map an inaccurate tracer of synchrotron emission at microwave frequencies.
(c) The spectral index map of $\beta(408\; {\rm MHz}, 23\; {\rm GHz}$) shows the 
flatter spectral 
index ($\beta \sim -2.5$) regions of active star-formation in the plane, where the cosmic ray 
electrons are generated.  The steeper spectral index regions ($\beta \sim -3$) off the 
plane suggest the energy losses suffered by the cosmic ray electrons during the period
of time required for their diffusion away from the star-formation regions of their origin.
This spectral index map is dominated by synchrotron emission, but still contains free-free 
emission.  It has been generated 
after setting zero-points based on cosecant fits to 
both maps, which provides an absolute zero-point for the \iMAP\ map. 
\label{haslamk}} 
\end{figure}

Figure \ref{haslamk}(a) shows the radio emission mapped over the
full sky with moderate sensitivity at 408 MHz by \citet{haslam/etal:1981}.  
This map is  
dominated by synchrotron emission.  Low frequency ($<
10$ GHz) spectral studies of the synchrotron emission indicate that $\beta \thickapprox
-2.7$, although substantial variations from this mean occur across the 
sky \citep{reich/reich:1988, lawson/etal:1987}.  
Higher energy cosmic ray electrons lose their energy faster than
lower energy electrons, so the synchrotron spectral index steepens with time.  Sometimes 
a ``break'' is seen in synchrotron spectra, which is useful for determining the age of the 
source.  At frequencies above the break the spectral index steepens to $\alpha=-(2\gamma +1)/3$ 
due to synchrotron losses.  This corresponds to an antenna temperature steepening of 
$\Delta\beta = -0.5$.  \citet{voelk:1989} predicts a break in the synchrotron spectrum of the 
Milky Way at 22 GHz, arising from a break in the relativistic electron spectrum at 20 GeV.
Synchrotron steepening breaks with $\Delta\beta \gt -0.5$ have been observed in a few discrete
sources 
\citep{green/scheuer:1992, morsi/reich:1987}, indicating additional processes beyond 
synchrotron losses.
 
Physical steepening effects compete against an observational measurement bias 
towards flat spectrum components.  Steep 
spectral index synchrotron components seen at low radio frequencies become weak relative 
to flat spectral index components when observed at higher microwave frequencies.  
Thus it should be expected that flatter spectrum components will increasingly dominate 
in observations at higher microwave frequencies, even as all components may steepen. 

The study of external galaxies can inform our picture of the Milky Way.  For example,
\citet{hummel/dahlem/vanderhulst:1991} present maps of the 610 MHz and 1.49 GHz emission 
from the edge-on spiral galaxy NGC891.  They find that the synchrotron spectral index
varies from $\beta\approx -2.6$ in most of the galactic plane to $\beta\approx -3.1$ 
in the halo.  Thus, it is reasonable to expect similar spectral index variations 
in the Milky Way at $\approx 1$ GHz.  
As one maps any such galaxy at increasingly higher frequencies 
the overall appearance will change considerably, with the flatter $\beta\approx -2.6$ 
plane emission increasingly dominating over the steeper spectral index halo.

In summary, {\it the synchrotron signal is complex.  A large range of spectral indices are both 
expected and observed. Thus, any synchrotron template map of the sky will be  
frequency-dependent}.  For many years CMB researchers have evaluated the synchrotron 
contamination level of their CMB data by examination of the Haslam-correlated amplitude 
scaled by $\beta \approx -2.75$.  Given the expected and observed spectral index 
variability, this not necessarily sound.

\subsection{Thermal Dust Emission and its Radio Correlation \label{thermdust}}

\begin{figure}
\figurenum{5}
\epsscale{0.5}
\plotone{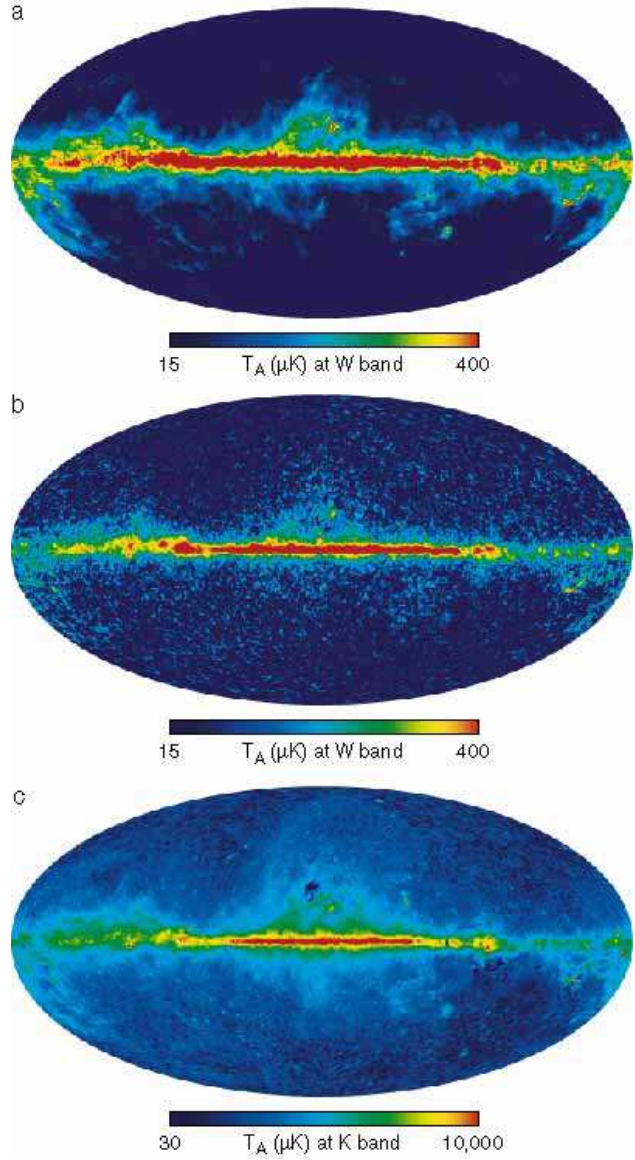}
\caption{(a) The FDS dust map at 94 GHz based on data 
from {\sl IRAS} and {\sl COBE}.  
This map is used as a prior in the MEM fit. \label{FDS}
(b) The full sky thermal dust map from the MEM procedure run on the five band \iMAP\ data and
shown for W-band.  
The morphology is found to be similar both to the expectation (prior), in (a) and to the 
synchrotron result map, in (c).  \label{MEMdust}
(c) This full sky map of synchrotron emission referred to 
K-band is based on the five bands of \iMAP\ 
measurements.  The map is the result of the MEM modeling process, as described in
the text.  Note the rough similarity of the microwave synchrotron emission to the thermal dust
emission.  This is presumed to be a result of their common origin in regions of 
star-formation. \label{MEMsync} 
} 
\end{figure}

Thermal dust emission has been mapped over the full sky in several infrared bands, 
most notably by the {\it COBE} and {\it IRAS} missions.  A full sky template 
is provided by \citet{schlegel/finkbeiner/davis:1998} (SFD), 
and is extrapolated in frequency by 
\citet{finkbeiner/davis/schlegel:1999}, hereafter referred to as FDS.  This is shown in 
Figure \ref{FDS}(a).

Measurements of the thermal dust spectral index generally lie in the range 
$1.5 \le \beta_d \le 2$.  For \iMAP\ wavelengths, the FDS extrapolation predicts 
$\beta_d\approx 1.6$.  \citet{schwartz:1982} showed that there is an inverse 
relationship between the dust temperature and its spectral index.  More recently, 
\citet{dupac/etal:2001, dupac/etal:2002} have observed this inverse relation, in detail, 
in sources observed with the PRONAOS (PROgramme NAtional d'Observations Submillimetriques)
2-m balloon telescope at effective wavelengths of 200, 260, 360 and 580 $\mu$m.  For 
dust temperatures below 20 K, 
\citet{dupac/etal:2002} observe values in the range $1.6\le \beta_d\le 2.5$ in M17.  
These steep
values have implications for the grain material.  They suggest emission from 
silicate grains such as MgO$\cdot$SiO$_2$ with $\beta_d\approx 2.5$ at $\approx 10$ K 
\citep{agladze/etal:1996} or natural olivines such as fayalite (Mg$_x$Fe$_{1-x}$SiO$_4$ 
with $x=0.06$) \citep{mennella/etal:1998}.

A remarkably tight correlation exists between the far-infrared and synchrotron emission 
of galaxies.  This relation has been extensively studied and modeled 
\citep{
dickey/salpeter:1984,
dejong/etal:1985,
helou/soifer/rowanrobinson:1985,
sanders/mirabel:1985,
gavazzi/cocito/vettolani:1986,
hummel:1986,
wunderlich/wielebinski/klein:1987,
wunderlich/klein:1988,
beck/golla:1988,
fitt/alexander/cox:1988,
hummel/etal:1988,
mirabel/sanders:1988,
bicay/helou/condon:1989,
devereux/eales:1989,
unger/etal:1989,
voelk:1989,
chi/wolfendale:1990,
wunderlich/klein:1991,
condon:1992,
bressan/silva/granato:2002}.

\cite{bicay/helou/condon:1989} comment that the synchrotron emission appears 
spatially smeared relative to the far-infrared distribution in galaxies.  They 
suggest that this likely arises from the diffusion and/or convection
of the cosmic ray electrons from their original site of star formation activity.  
If this is true, then one would expect higher frequency radio observations to become 
increasingly tightly correlated with the far-infrared distribution since the 
higher microwave frequencies preferentially sample the flatter spectrum, 
younger component.

Commenting on the synchrotron-dust correlation, \citet{voelk:1989} predicts that, 
``...whether integral or local, the correlation should also 
become better for smaller wavelengths of observation.  The reason is that a decreasing 
wavelength implies higher energies of the radiating electrons.  These electrons lose 
energy faster and therefore are more localized around their sources.''

\subsection{Spinning and Magnetic Dust Emission}

In addition to the thermal emission from dust, above, there are various other ways in which dust 
can radiate at microwave wavelengths.  These include electric dipole emission from spinning dust
grains and magnetic dipole emission from thermally fluctuating dust grains 
\citep{erickson:1957,  
draine/lazarian:1999, draine/lazarian:1998a, draine/lazarian:1998b}.  
Emission from spinning dust can produce $\beta \sim  -2$ from 20-40 GHz.   The high-frequency 
cut-off of spinning dust emission is due to the limited speed at which a dust grain can spin.

All theories to explain the amazingly tight correlation between radio and far-infrared 
emission from galaxy to galaxy relate to the level of the star formation activity.  
The star formation cycle forms and destroys dust grains, heats the grains, creates magnetic 
fields and relativistic electrons, and creates the O- and B-stars that ionize the gas.  
The tight correlation persists down to very low radio frequencies, yet the radio emission can not 
be due to spinning dust gains since they do not radiate strongly at such long wavelengths. This 
suggests that radio/microwave emission and thermal dust emission are tightly 
correlated, not because of a common dust emission source, but rather as a result of the fact that 
both trace star-formation activity.  While spinning dust emission is not a dominant source of 
$\sim 1$ GHz radio emission in the Galaxy, it may be significant at higher microwave frequencies
($\sim 30$ GHz).

\citet{finkbeiner/etal:2002} searched for spinning dust emission in ten discrete 
Galactic sources at 5, 8, and 10 GHz.
No detection was reported for 8 of the 10 sources, implying that spinning dust emission 
is not a ubiquitous and dominant emission mechanism in our Galaxy.  
Two of the ten sources were reported 
``tentative'' detections.  They caution that a solid detection would require 
observations of the predicted spectral roll off both at high and low frequencies.  

It is reasonable to assume that the Milky Way should be like other spiral galaxies and, in
particular, the microwave properties of external galaxies should help elucidate the global 
properties of the Milky Way.  
\citet{klein/emerson:1981} report the spectra of 16 spiral galaxies to frequencies up to 
10.7 GHz.  They find that the spectral index, averaged over the 16 galaxies, is well 
described by a single synchrotron power law spectral index of $\beta=-2.71\pm 0.08$.  
\citet{gioia/gregorini/klein:1982} report that the mean spectrum of 56 bright
spiral galaxies is dominated by synchrotron emission, with a small additional free-free 
emission component.  They report that a free-free plus single power-law synchrotron fit 
fully explains their observations.  \citet{niklas/klein/wielebinski:1997} examine 
74 Shapley-Ames galaxies and they also find their 
spectra dominated by a power-law synchrotron index with some free-free emission.
These extragalactic studies report no suggestion of anomalous sources of emission.
The main impediment to stronger statements about spinning dust emission in external 
galaxies is the (perhaps surprising) sparsity of microwave observations of normal 
galaxies at microwave frequencies greater than $\sim 10$ GHz.  

\section{INTERNAL LINEAR COMBINATION OF THE MULTIFREQUENCY \iMAP\ MAPS \label{ilc}}

Linear combinations of the multi-frequency \iMAP\ sky maps can be formed using 
constraints so that the response to Galactic signals is approximately canceled
while maintaining unit response to the CMB.
Linear combinations of multi-frequency data need not rely on assumptions about
the foreground signal strengths of the various emission mechanisms.  
An advantage of this method is that it is ``internal'', that is, it relies only on
\iMAP\ data, so the calibration and systematic errors of other experiments do not enter.

We create a CMB map by computing a weighted combination of the maps that have been
band-averaged within each of the five \iMAP\ frequency bands, 
all smoothed to $1^\circ$ resolution.  
Weights can be determined by minimizing the variance of the measured temperatures 
subject to the constraint that the weights add to one, thus preserving unit 
response to the thermal CMB spectrum.

When the entire sky is treated with a single set of 
weights, imperfections are apparent due to the fact that the foreground properties 
(i.e., spectral indices) have significant spatial variation, especially within the 
inner Galactic plane.  Thus, to improve the result, the inner Galactic plane is  
divided into 11 separate regions, within which weights are determined independently.
All but one of the $N=12$ total regions are thus in the inner Galactic plane, 
within the Kp2 cut.  
The temperature (in CMB thermodynamic units) of each pixel, $p$, in each region  
$n=1, 2, ...N$ is
\begin{equation}
T_n(p) = \sum^5_{i=1} \zeta_{n,i} T^i(p),
\end{equation}

\noindent where the sum is over the five \iMAP\ bands.  
The weights, $\zeta$, for each region are determined by carrying out a nonlinear 
search for the minimum of $\mathrm{var}(T_n)$ within each of the $N$ regions of the sky, 
subject to the constraint that 
\begin{equation}
\sum^5_{i=1} \zeta_i = 1,
\end{equation}
thus preserving only signals with a CMB spectrum. 
For the region that covers the full sky outside of the inner Galactic plane, 
the weights are $\zeta = 
 0.109,   -0.684,   -0.096,   1.921,   -0.250
$ for K, Ka, Q, V, and W bands, 
respectively.

\begin{figure}
\figurenum{6}
\epsscale{0.5}
\plotone{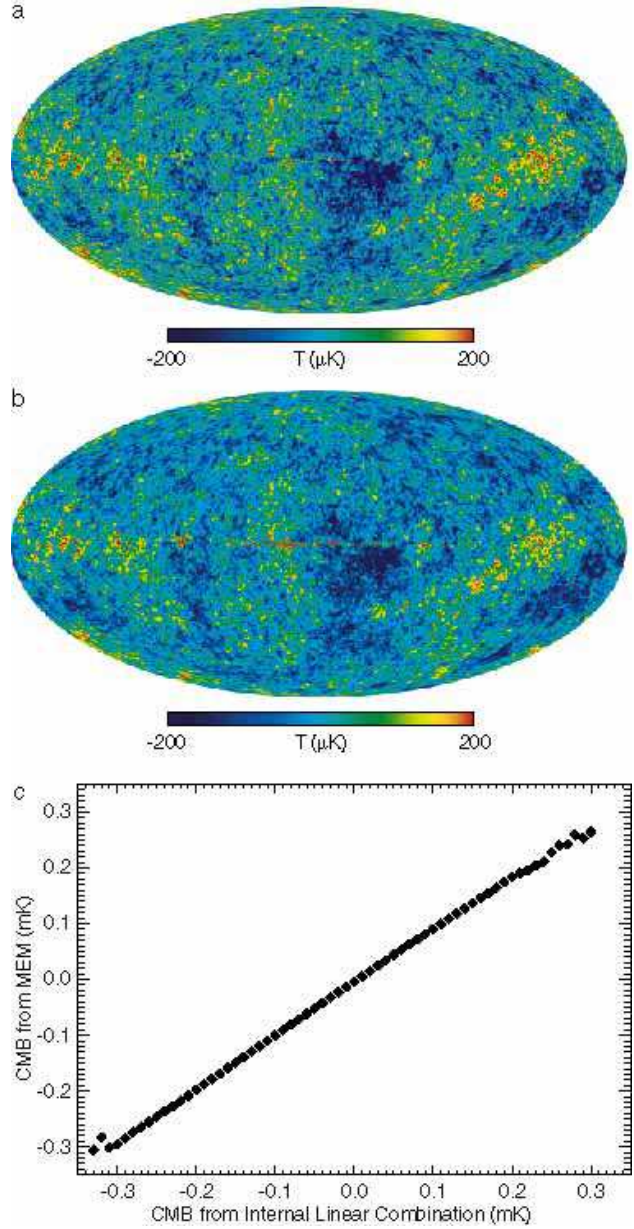}
\caption{
(a) A \iMAP\ CMB full sky map derived from an internal combination of the \iMAP\ data 
with no use of external data sets, as described in \S\ref{ilc}. \label{bs}
(b) The full sky CMB map that results from the MEM procedure, as described in \S\ref{mem}. 
\label{MEMcmb}
(c) The plot is a binned pixel-by-pixel comparison of the MEM-derived CMB map with the 
CMB map derived by the minimum variance combination technique.
\label{TTMEMCMB}
} 
\end{figure}

This process creates a map with slight discontinuities at the region boundaries 
since each region, $R_n$, is treated independently.  To smooth these discontinuities, we 
create a set of $N$ maps, with weights $w_n(p)=1$ for $p\in R_n$ and $w_n(p)=0$ 
for all other $p$.  We then smooth these $N=12$ maps of $w_n(p)$ (containing only 
ones and zeros) using a $1\ddeg5$ smoothing kernel.  The final map is then given 
by the weighted combination  
\begin{equation}
T(p)=\sum_{n=1}^N w_n(p)T_n(p)\; \bigg/ \; \sum_{n=1}^N w_n(p),
\end{equation}
which greatly reduces boundary effects.  The resulting CMB map is shown in Figure 
\ref{bs}(a).

\section{MAXIMUM ENTROPY METHOD (MEM) SPATIAL AND SPECTRAL FIT\label{fitting} \label{mem}}

In \S{\ref{ilc}} the objective was 
to cancel the Galactic signal, regardless of its nature.  
In this section we attempt to distinguish and understand the 
different emission sources.  
It is a coincidence of nature that there are three known emission mechanisms that are 
each capable of generating similar spectra in the microwave spectral regime.  Free-free, 
synchrotron, and spinning dust emission can all have a spectral index of
$\beta \sim -2$  at the \iMAP\ frequencies.  
The key to separating these is to use data from outside the microwave spectral regime.  

Recent observations of H$\alpha$ emission over large regions of the sky can help 
trace free-free emission.  This is especially useful in regions 
where the extinction of H$\alpha$ is small, largely at high Galactic latitudes.
The 408 MHz radio map \citep{haslam/etal:1981} is at a sufficiently low frequency 
that spinning dust emission is negligible.  

We formulate a model temperature solution at each frequency band and each pixel, $p$, as
\begin{equation}
T_m(\nu, p) = S_{cmb}(\nu \vert p)T_{cmb}(p) + S_s(\nu\vert p) T_s(p) + 
S_{ff}(\nu\vert p)T_{ff}(p) + S_d(\nu\vert p) T_d(p)
\end{equation}

\noindent where $cmb, s, ff,$ and $d$ represent the CMB, synchrotron (plus spinning dust), 
free-free, and thermal dust components,
respectively.  $S_c(\nu\vert p)$ is the spectrum of emission $c$ in pixel p, and is not
necessarily uniform across the sky.  $T_c (p)$ is the spatial distribution of emission type 
$c$, where synchrotron (plus spinning dust) 
and free-free are normalized to the \iMAP\ K-band temperature, and thermal dust is
normalized at W-band.  The model is fit by minimizing the functional
$H(p)=A(p) + \lambda (p) B(p)$ \citep{press/etal:NRIC:2e}, 
where $\lambda$ is a regularizing parameter, 
$A(p) = \chi^2(p) = \sum_\nu[T(\nu,p)-T_m(\nu,p)]^2/\sigma^2$ 
is a pixel by pixel fit, and 
$B(p)=\sum_c T_c (p) \ln [ T_c(p)/P_c (p) ]$ 
sums only over Galactic emission components.  $P_c (p)$ is a prior model of the spatial 
distribution of the temperature of emission component, $c$, normalized to the same frequency 
as $T_c$.  The MEM parameter $\lambda$ controls
the fidelity of the fit to the a priori model.  The form of $B$ ensures 
the positivity of the Galactic 
emission $T_c (p)$, which greatly alleviates degeneracy between the foreground components.
The formalism is unaware, for example, of the different physics of synchrotron versus 
spinning dust emission.

\begin{figure}
\figurenum{7}
\epsscale{1}
\plotone{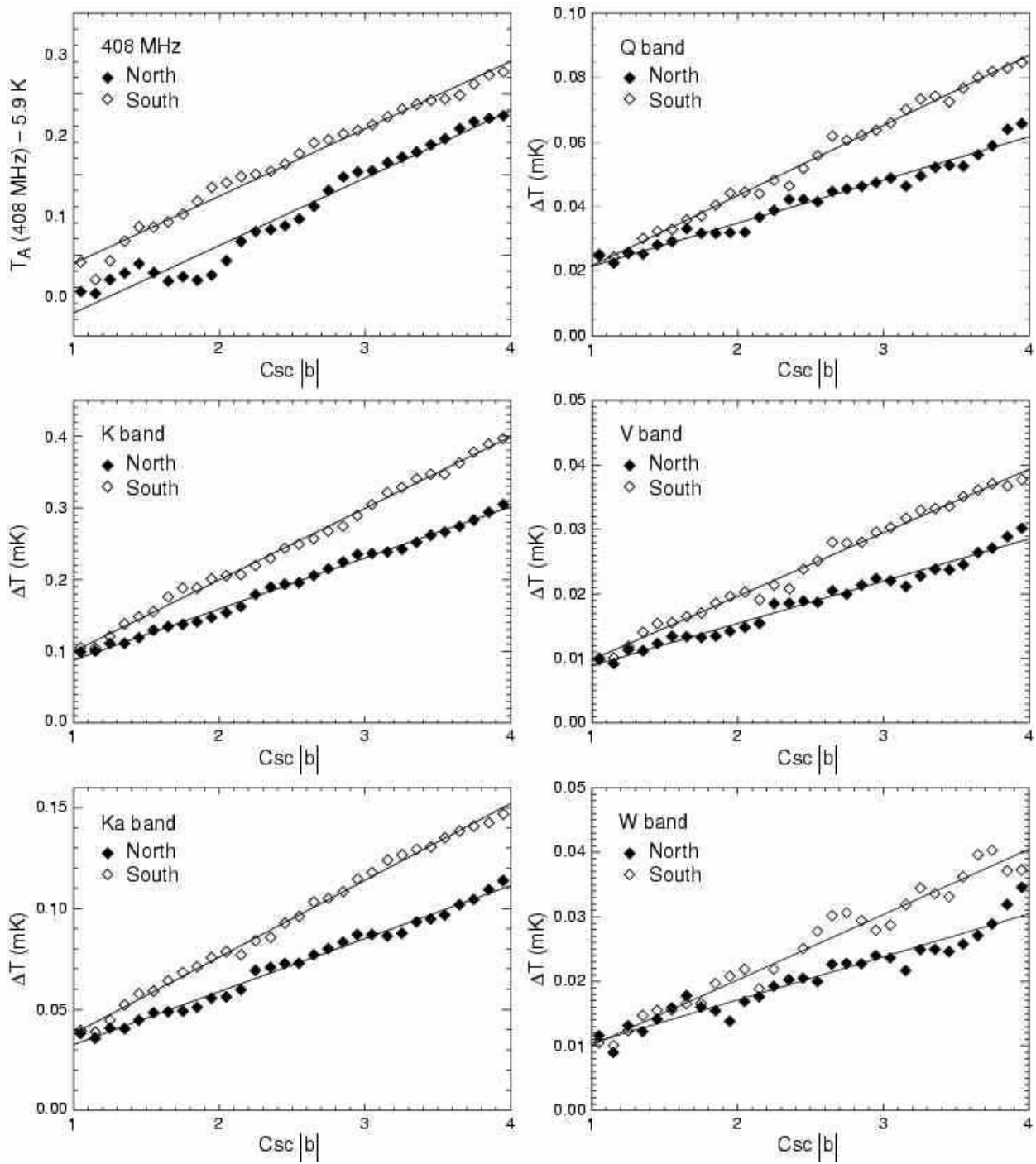}
\caption{
\footnotesize
A plane-parallel cosecant model gives a fairly good fit to the 408 MHz Haslam map
and the five frequency \iMAP\ maps.  Application of a cosecant fit is used to set 
the zero-points of the \iMAP\ maps in a self-consistent manner.  The Southern fit  
is used to minimize the number of negative map temperatures.  Each panel shows linear fits 
to data averaged over bins in csc$(b)$ for both Northern and Southern Galactic latitudes.
For the Northern latitudes, the longitude range covering the North Polar Spur was excluded.
A 5.9 K extragalactic correction \citep{lawson/etal:1987} was subtracted from the 408 MHz 
data, and the \iMAP\ data shown here were normalized such that the intercept of the csc$(b)$ 
fit for Southern latitudes is zero for each band.
\label{cscplots}} 
\end{figure}

We begin by smoothing all maps to a uniform $1^\circ$ resolution.  
Initially we subtract the internal 
combination CMB map derived in \S\ref{ilc} from the five \iMAP\ maps to remove this 
part from the fit.  
Since \iMAP\ measures only differentially on the sky it does not determine a meaningful  
zero level for the maps.  
For each of the 5 maps, we use the mean observed temperature variations with Galactic 
latitude to estimate the zero level, assuming a plane-parallel model for the Galactic 
emission.  This is done by adding an offset to each map such that a fit of the form 
$T=A\;{\rm csc}\;b+B$ over the range $-90^\circ < b < -15^\circ$ yields a value of zero for the
intercept, $B$.  For consistency we treat the 408 MHz map the same way.  Figure \ref{cscplots}
shows fits of this form for $b<-15^\circ$ and for $b>15^\circ$.  The results for 
the Southern sky are chosen because they minimize the 
number of negative pixel values in the 408 MHz map.

The prior spatial distribution maps, $P_c(p)$, are chosen as follows.  
For synchrotron emission we use the Haslam map scaled to K-band using  
$\beta_s=-2.9$.  For free-free, we use the 
full sky H$\alpha$ map produced by Finkbeiner, scaled to K-band antenna 
temperature using 11.4 $\mu$K R$^{-1}$.  The dust prior is taken to be 
``Model 8'' of FDS, evaluated at 94 GHz.  These priors are fixed, but 
we have evaluated the sensitivity of the output solution to the choice 
of prior and, as discussed below, have used the results to guide the 
choice of the regularizing parameter $\lambda$.

We construct initial spectral models, $S_c(\nu \vert p)$, as follows. 
From the 408 MHz map (Figure \ref{haslamk}a) and the \iMAP\ K-band 
map (Figure \ref{haslamk}b)
we construct the $\beta(408\;{\rm MHz},\; 23\;{\rm GHz})$ spectral index map in
Figure \ref{haslamk}c.  This is largely a synchrotron spectral index map, but with
some free-free and spinning dust contribution.
Significant variations are seen in this spectral index across the sky.  The initial
model for $S_s(\nu \vert p)$ is $(\nu/\nu_K)^{\beta(p)-0.2}$
where $\beta(p)$ is the above spectral index map.  For the free-free and dust, we
take the initial models to be 
$S_{ff}(\nu) = (\nu/\nu_K)^{-2.15}$ and $S_d(\nu) = (\nu/\nu_W)^{+2.2}$ 
respectively. The latter choice for the dust model was informed by many 
previous trial models.

In general, for a given set of spectral models, the MEM solution has 
residual errors.  Since the synchrotron and dust spectra are poorly known 
at microwave frequencies, we use the model residuals to guide an update of 
the synchrotron and dust spectral models, $S_s$ and $S_d$ in an iterative 
fashion.  Specifically, after each minimization of $H(p)$, we update $S_s$ 
and $S_d$ as follows
\begin{eqnarray}
T_s S_s(\nu \vert p) \rightarrow T_s S_s(\nu \vert p) + g\cdot R(\nu,p) & & {\rm K-V\; bands} \\
T_d S_d(\nu \vert p) \rightarrow T_d S_d(\nu \vert p) + g\cdot R(\nu,p) & & {\rm W\; band}
\end{eqnarray}
where $R(\nu,p) = T(\nu,p)-T_m(\nu,p)$ is the solution residual (data minus model), 
and $g$ is 
a gain factor ($g\le 1$).  Each spectral model is renormalized 
to 1 at its fiducial frequency prior to running a new minimization of $H(p)$.  
To stabilize the MEM computations 
at frequencies where individual components are weak, and thus poorly determined, 
we extend the synchrotron power law spectral index between Q-band and V-band to 
W-band.  Likewise, we extend the dust index between W- and V-bands to Q-, Ka-, 
and K-bands.  The extension is re-applied at every iterative step.    

The regularizing parameter, $\lambda$, controls the degree to which the solution 
follows the prior.  Note that $\lambda(p)$ can vary across the sky since the fits are performed 
independently for each pixel.  Our philosophy is that the \iMAP\ data should take priority over 
the priors in sky regions where the \iMAP\ signal-to-noise ratio is high and that the priors 
should take priority where the \iMAP\ signal-to-noise ratios are too low.  
After some trial-and-error, we arrive 
at a map of $\lambda(p)$ values that implement this, 
$\lambda(p)=0.6\;(T/1\;{\rm mK})^{-1.5}$, where $T$ is the K-band antenna 
temperature in mK.  This choice makes $\lambda(p)$ vary continuously between $\sim 3$ where 
the K-band signal is noisy and $\sim 0.2$ where the K-band signal is strong.  The results are not
strongly sensitive to the exact form of $\lambda(p)$.  Similar results were obtained with
$\lambda=0.2$ everywhere, but the results are noisy where the \iMAP\ signal is weak.  Using
$\lambda>3$ everywhere produces results that are somewhat dependent on the priors, $P$, in 
regions of high signal-to-noise ratio.

\begin{figure}
\figurenum{8}
\epsscale{0.5}
\plotone{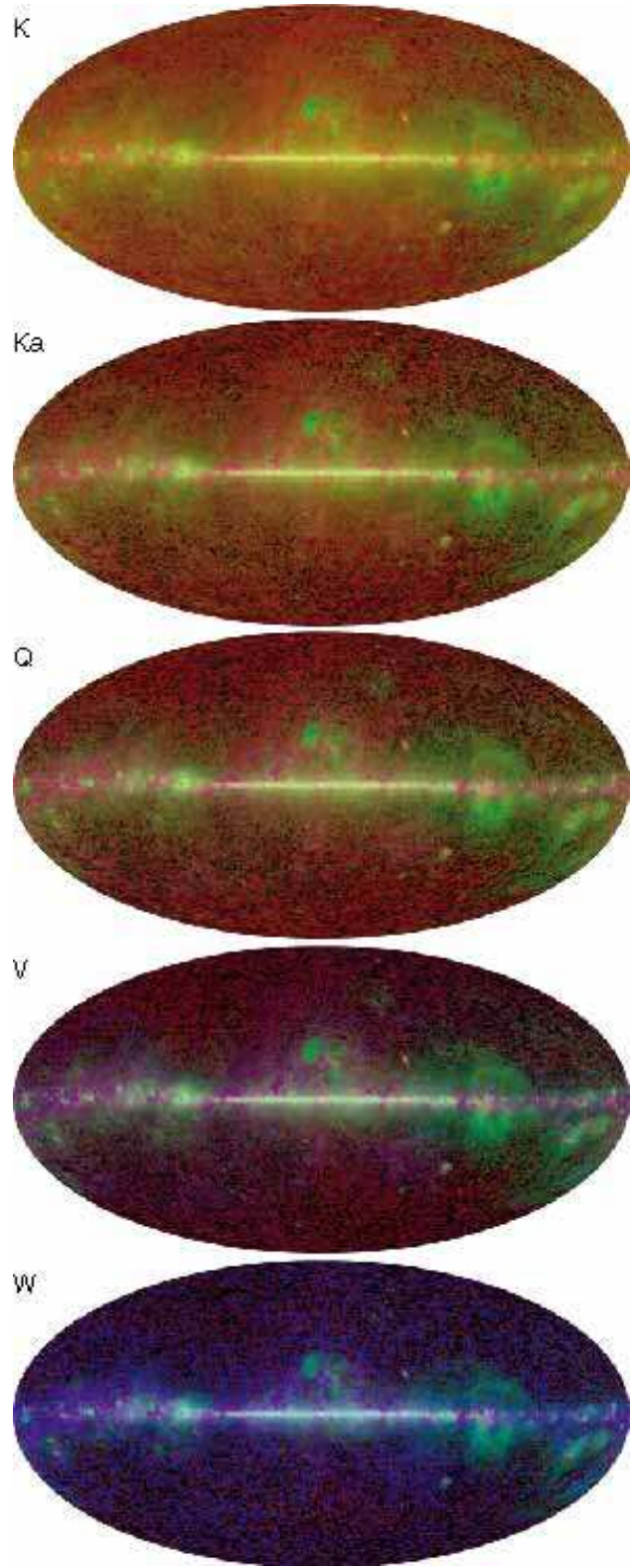}
\caption{
\footnotesize
Three color maps of the Galactic emission from the the MEM model for K-band (top) 
through W-band (bottom).  
These maps give a feeling for which emission mechanism dominates as a function of frequency and sky
position.  Synchrotron is red, free-free is green, and the thermal dust is blue.  
\label{3color}
} 
\end{figure}

\begin{figure}
\figurenum{9}
\epsscale{0.7}
\plotone{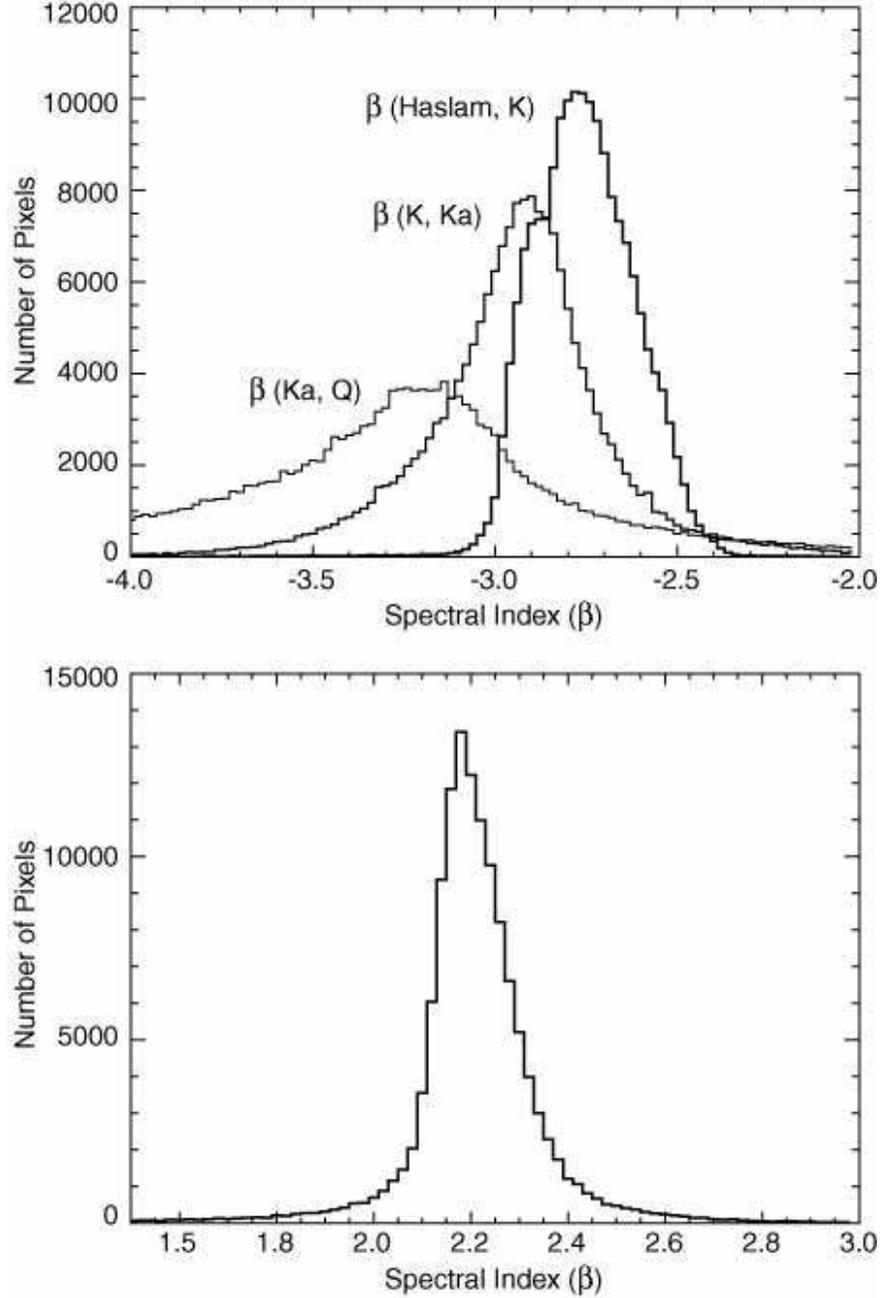}
\caption{
\footnotesize
{\it Top}:  
Spectral index histograms of the MEM-fit low frequency component.  Lower amplitudes 
and increased widths of distributions are indicative of lower signal-to-noise ratios rather than 
physical effects.  The histogram only includes pixel values where the total K-band antenna 
temperature is T$_A$(K-band)$>500\mu$K to minimize low signal-to-noise artifacts. 
The steep spectral 
index $\beta\sim -2.75$ between the Haslam 408 MHz measurements and the \iMAP\ MEM solution 
at K-band is seen to steepen further in going from K-band to Ka-band and even further still 
to Q-band.  
The total steepening 
of $\Delta\beta\approx -0.5$ is highly suggestive of synchrotron emission with a synchrotron 
loss spectral break near K-band.
There is no evidence for the spectral flattening expected from spinning dust emission.  We can 
thus limit spinning dust emission to $<5$\%
of the overall Ka-band signal.\label{synchist}
{\it Bottom}: Spectral index histogram of the MEM-fit dust component. The MEM fit to thermal 
emission from dust gives a spectral index of $\beta_d\approx 2.2\pm 0.1$ between V-band and 
W-band.  The wings of the
distribution are dominated by signal-to-noise limitations.\label{dusthist}
} 
\end{figure}

We run the MEM code to minimize $H$.
We typically iterate (residual errors fed back to improve the solutions) $\sim 10$ times with 
a feedback gain factor of $g=0.5$ per iteration to produce overall residuals $<1$\%.
The results of the MEM are shown in Figures \ref{MEMff}, \ref{MEMdust}, and \ref{3color}.
Note that minimal residuals and reasonable results are achieved without the introduction 
of any unmodeled Galactic component.  Figure \ref{MEMsync} 
shows that 
there is a fairly close spatial correlation between synchrotron and thermal dust emission
at microwave wavelengths.  Figure \ref{synchist} is a histogram of the 
synchrotron and/or spinning dust component.  The spectral index looks very much like 
synchrotron emission with a 0.5 steepening of the index from synchrotron losses, as 
predicted by \citet{voelk:1989} based on the observed break in the relativistic 
electron spectrum in the solar neighborhood. 

To limit the fraction of the low frequency non-thermal component that can be attributed 
to spinning dust, we construct a spinning dust emission model whose spectrum is 
characterized by the Cold Neutral Medium spectrum of \citet{draine/lazarian:1998a}, and 
whose spatial distribution is that of the MEM-derived thermal dust emission.  The 
overall normalization of the model is a free parameter.  For a given normalization factor, 
we subtract this map from the non-thermal component model maps at each of the five WMAP 
bands and at the Haslam frequency; the remainder is assumed to be a series of synchrotron 
maps.  We compute the median values of $\beta$(408 MHz, K-band), $\beta$(K-band, Ka-band), 
and $\beta$(Ka-band, Q-band) for the residual maps and demand that the residual 
(synchrotron) spectral index be a monotonically decreasing function of frequency in the 
range 408 MHz to 41 GHz (Q band). The result is that the spinning dust
fraction is less 
than about 5\% of the total Ka-band emission.  We conclude that any spinning dust 
emission component is subdominant.   The thermal dust spectral index histogram, 
shown in Figure \ref{dusthist}, indicates $\beta_d\approx 2.2\pm 0.1$.

\begin{figure}
\figurenum{10}
\epsscale{0.5}
\plotone{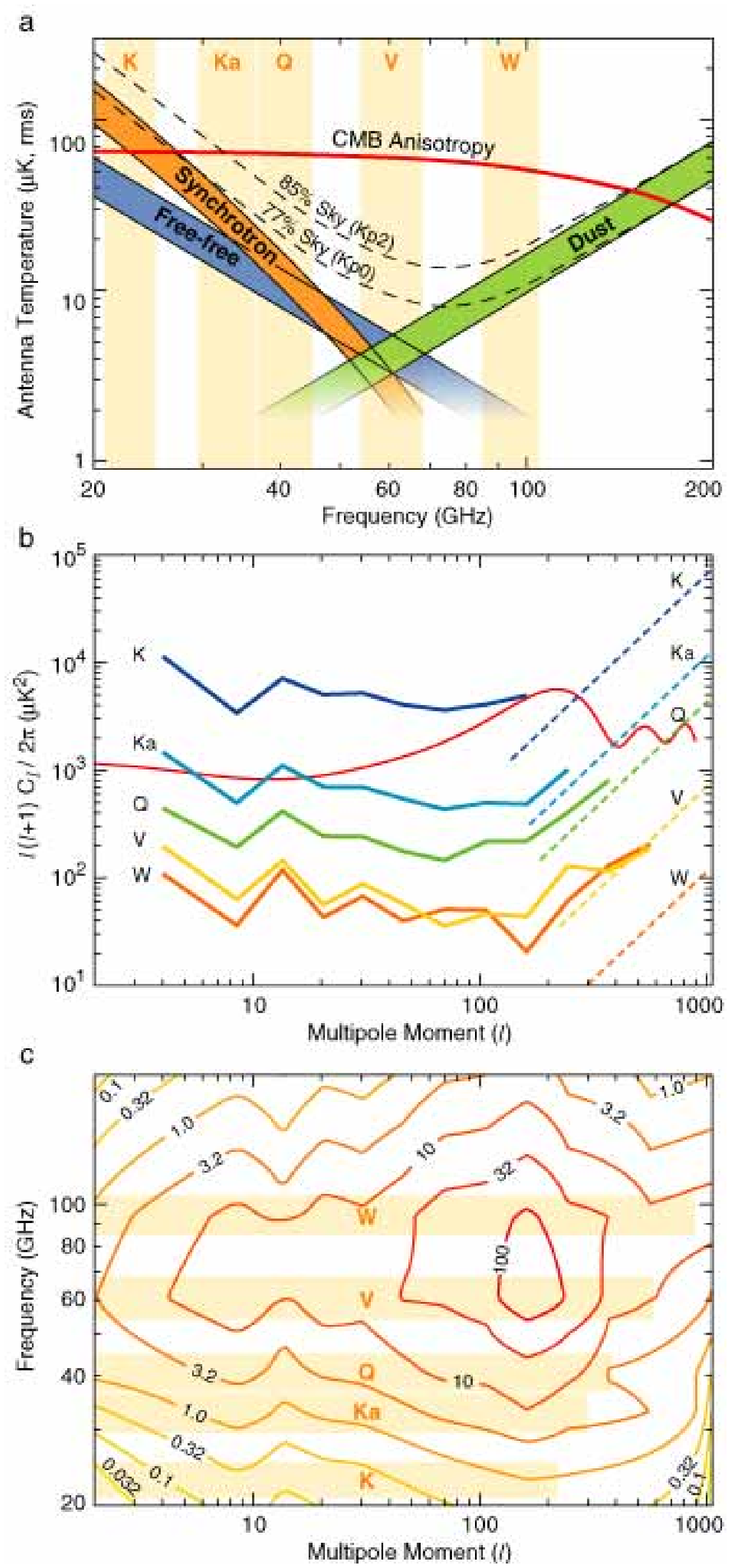}
\caption{
\footnotesize
CMB versus Foreground Anisotropy.  
The \iMAP\ frequency bands were chosen to be in a spectral region where the CMB anisotropy 
is most dominant over the competing Galactic and extragalactic foreground emission. This 
can now be quantified for \iMAP\ and future experiments.
(a) Spectra are shown of the CMB anisotropy and the Galactic emission from the MEM modeling.
(b) The foreground power spectra are shown for each \iMAP\ band using the Kp2 mask.  
The power spectra within frequency bands
are derived as cross-power spectra between radiometers minus the \iMAP-derived CMB model.  
In K- and Ka-band, where cross-power spectra are not available, the noise bias has been
estimated and subtracted.  The band-by-band point source fits to an $l^2$ term are shown 
in the dashed 
lines on the right.  The power spectra are expected to asymptotically join these lines.
Note that the total foreground power spectra (excluding point sources) go as 
$C_l\sim l^{-2}$.
(c) The contour plot shows the ratio of CMB to foreground anisotropy power as a function 
of frequency and multipole moment for the Kp2 mask.
\label{overview}}
\end{figure}

The MEM method is a Bayesian approach to model fitting that incorporates prior 
knowledge of the Galactic model.  The results are driven by a
combination of the data and the priors. The CMB internal linear combination 
map is based on 5 numbers over the majority of the sky, therefore subtracting it 
from each band removes only a small fraction of a degree of freedom per pixel. Then 
we fit 7 parameters for free-free, synchrotron (+spinning dust), and thermal dust: 
3 amplitudes, and 4 spectral indices (0 free-free, 3 synchrotron, 1 thermal dust).  
We fit these 7 parameters with 5 bands of data plus 3 amplitude priors 
(which incorporate positivity on each amplitude), for a total of 7
parameters with 8 independent 
constraints per pixel.  The degree to which the prior information constrains the 
model is set by the regularizing parameter $\lambda$, thus one cannot count the 
number of degrees of freedom as in a pure $\chi^2$ fit.
A demonstration that there is sufficient power in the data to guide the fit is that 
the result is relatively insensitive to variations in the priors or the gain and 
that the results are physically reasonable.

We freeze the spectral models $S_s$, $S_{ff}$, and $S_d$ once small residuals 
are achieved, and 
the improvement in the residuals is not significant.  We then introduce the CMB 
as a model component with a known spectral index, 
$S_{cmb}(\nu\vert p) = dB/dT$ for a Planck function $B$ with $T_0=2.725$ K, 
and iterate again to solve for a CMB map and modified component amplitude maps.
The MEM-derived CMB map is shown in Figure \ref{MEMcmb}b and a binned pixel-by-pixel comparison
with the internal combination CMB map is shown in Figure \ref{TTMEMCMB}c.
It is not surprising that the MEM CMB map agrees well with the linear combination map 
since the foreground components were derived based on low residuals with respect to the 
subtracted linear combination CMB map.  Yet, by giving the component amplitudes an opportunity to 
re-adjust with the CMB amplitudes as a new degree of freedom, it is a reassuring cross-check 
that the solutions are generally unaffected.  Small ($\approx 1$\%) errors seen in the 
Galactic plane of the MEM CMB map relative to the combination CMB map illustrate the 
residual uncertainty of the MEM solution.

Low residual solutions are highly constrained, but not necessarily unique or correct.  
Loosely speaking, the model must account for a 
considerable signal amplitude in K-band and very little total amplitude in V-band.  This 
constrains the synchrotron spectral index to be steep.  Likewise, the model must match 
the dust signal in W-band, but the dust spectral index must be steep enough to stay under 
the total signal level 
in V-band.  The free-free level can be thought of as making up the remainder of the V-band 
signal after the synchrotron and dust have been fit.   
Any additional emission components are strongly disfavored due to the already low total V-band
signal.
Limitations in the precision and accuracy of the H$\alpha$ measurements, uncertainty in the 
degree of H$\alpha$ extinction, and the 
uncertainty in the conversion factor between H$\alpha$ and free-free, combine so as to not 
allow the 
free-free to be fixed with sufficient precision to achieve a uniquely ``correct'' MEM 
solution.  
Figure \ref{TTMEMff}c shows a (bin-averaged) pixel-by-pixel comparison of the 
MEM free-free result with the 
H$\alpha$ prior.  The slope is within $\pm12$\% (depending on the method of fit) 
of the expected value from \S\ref{ff} and Table \ref{tbl-1}, 
meaning that the model can not be far off what is physically reasonable.

The total observed Galactic emission matches the MEM model to $<1$\%
while emission is separated into the individual components with a few percent accuracy.
An overview of the Galactic spectra and spatial power is shown in Figure \ref{overview}.
The total flux of our Galaxy ($F_\nu = \int I_\nu(l,b) \cos l \cos b \; d\Omega$) 
in the five \iMAP\ bands is 45, 38, 35, 32, and  
48 kJy, from K-band to W-band.
Thus, we present a ``reasonable'' model that conveys far more information 
about the microwave foregrounds than was previously known.

\section{TEMPLATE FITS \label{templatefits}}

While the MEM method, discussed above in \S\ref{mem}, produces a CMB map, 
it is not straightforward to use that map for
cosmological analyses since its noise properties are complicated.  The MEM method is much more
valuable for understanding the physical nature of the foregrounds than for 
producing a CMB map useful for CMB analysis.

To produce a CMB map useful for cosmological analyses (by maintaining well specified noise
properties) we simultaneously fit (outside the Kp2 cut) 
a set of externally derived template maps to the \iMAP\
maps.  The template maps are the H$\alpha$ map (corrected for an estimate of 
extinction), the 408 MHz Haslam map, 
and the FDS 94 GHz dust map.  The resulting coefficients are given in Table \ref{tbl-templates}.
A constant term is included in the fits.   

\begin{deluxetable}{ccccccc}
\tablecaption{Template Fits to the \iMAP\ Maps:  Coefficients and Percent 
Fit by External Template 
\label{tbl-templates}}
\tablewidth{0pt}
\tablehead{
\colhead{Band} &  \colhead{Dust:FDS}  &  \colhead{Free-Free:H$\alpha$} &
\colhead{Synchrotron:Haslam} &
FDS\tablenotemark{a}  &  H$\alpha$\tablenotemark{a}  &  Haslam\tablenotemark{a}  \\
& \colhead{(rel to FDS)} &  \colhead{($\mu$K/R)}  & \colhead{($\mu$K/K)} &
(\%) & (\%) & (\%)
}
\startdata
K    &        6.3       &   4.6       &  5.6  &  60  & 14  & 26 \\
Ka   &        2.4       &   2.1       &  1.5  &  64  & 18  & 18 \\
Q1   &        1.5       &   1.3       &  0.5  &  69  & 20  & 11 \\
Q2   &        1.4       &   1.3       &  0.5  &  68  & 20  & 12 \\
V1   &        0.9       &   0.5       & -0.2  &  92  & 15  & $-$8 \\
V2   &        0.9       &   0.4       & -0.2  &  95  & 13  & $-$8 \\
W1   &        1.2       &   0.1       & -0.3  & 113  &  3  & $-$16\\
W2   &        1.2       &   0.1       & -0.4  & 113  &  4  & $-$17\\
W3   &        1.2       &   0.1       & -0.3  & 112  &  2  & $-$14\\
W4   &        1.1       &   0.1       & -0.3  & 112  &  4  & $-$16\\
\enddata
\tablenotetext{a}{The percentages are estimates of how the total fit signal 
breaks down into the three templates, for each frequency band.}
\end{deluxetable}

\begin{deluxetable}{ccccc}
\tablecaption{\iMAP\ Point Source Counts
\label{tbl:src:dnds}}
\tablewidth{0pt}
\tablehead{ Band & Flux Range & $N_{\rm src}$ &      $\kappa$         & $\beta$ \\
            &   (Jy)     &               & (Jy$^{-1}$ sr$^{-1}$) &
}
\startdata
 K  & 2-10 & 72 & $45\pm12$ & $-2.8\pm0.2$ \\
 Ka & 2-10 & 73 & $44\pm12$ & $-2.8\pm0.2$ \\
 Q  & 2-10 & 61 & $32\pm9$ & $-2.7\pm0.2$ \\
\enddata
\end{deluxetable}

The rising amplitude of the dust template fit coefficient with 
decreasing frequency has been interpreted by some as evidence of spinning dust emission.  
We have learned from the MEM solutions in \S\ref{mem}, however, that the 408 MHz Haslam map is 
not a high-fidelity synchrotron template at the \iMAP\ frequencies due to the strongly 
variable spectral index across the sky.  This is reflected in 
the negative template coefficients in V- and W-bands.  From the point of view of removing
foreground emission for CMB analyses, we are not concerned about the degeneracy that exists
between the strongly correlated dust and synchrotron emission components.  We are only concerned
with reducing the foreground level.

To further investigate past results on dust-correlated radio emission we fit the FDS 
thermal dust template to the total MEM 
Galactic model (synchrotron plus free-free plus thermal dust emission) 
for each \iMAP\ band, using the Kp2 cut.  We then compute the spectral index of the dust-correlated
emission.  We use the \iMAP\ Ka and V bands to approximate the {\sl COBE} 
DMR 31 GHz and 53 GHz bands.  
The result is $\beta$(31 GHz, 53 GHz) $\approx \beta$(Ka, V)$ = -2.2$.  This replicates the 
previous dust-correlated radio emission.  Physically, the $-2.2$ index is a combination of a
$\beta\approx -3$ synchrotron component that is correlated with thermal dust (because of 
star-formation activity) combined with the $\beta=+2.2$ thermal dust component.  
This combined index gets much steeper at lower frequencies (where the dust emission 
is weak) and is much flatter at higher frequencies where the synchrotron and free-free  
become weak relative to thermal dust emission.

An alternative fitting method is used in the current first-year \iMAP\ 
CMB analyses to avoid negative coefficients for the synchrotron
component, above.  Only the Q-, V-, and W-band data are included.  The dust
component is fitted separately to each band (3 parameters).  A single
amplitude is used to fit the free-free emission across all three bands,
assuming a spectral index $\beta=-2.15$.  Similarly, a single amplitude is
used for the synchrotron emission, $\beta=-2.7$.  Thus, five
parameters are used:  dust amplitudes in the Q-, V-, and W-bands (1.036, 0.619, 0.873, 
respectively), a free-free amplitude (1.923), and a
synchrotron amplitude (1.006).  The units of the coefficients are the same
as in Table \ref{tbl-templates}, with the free-free and synchrotron amplitudes referred to 
Q-band.

\begin{figure}
\figurenum{11}
\epsscale{1.0}
\plotone{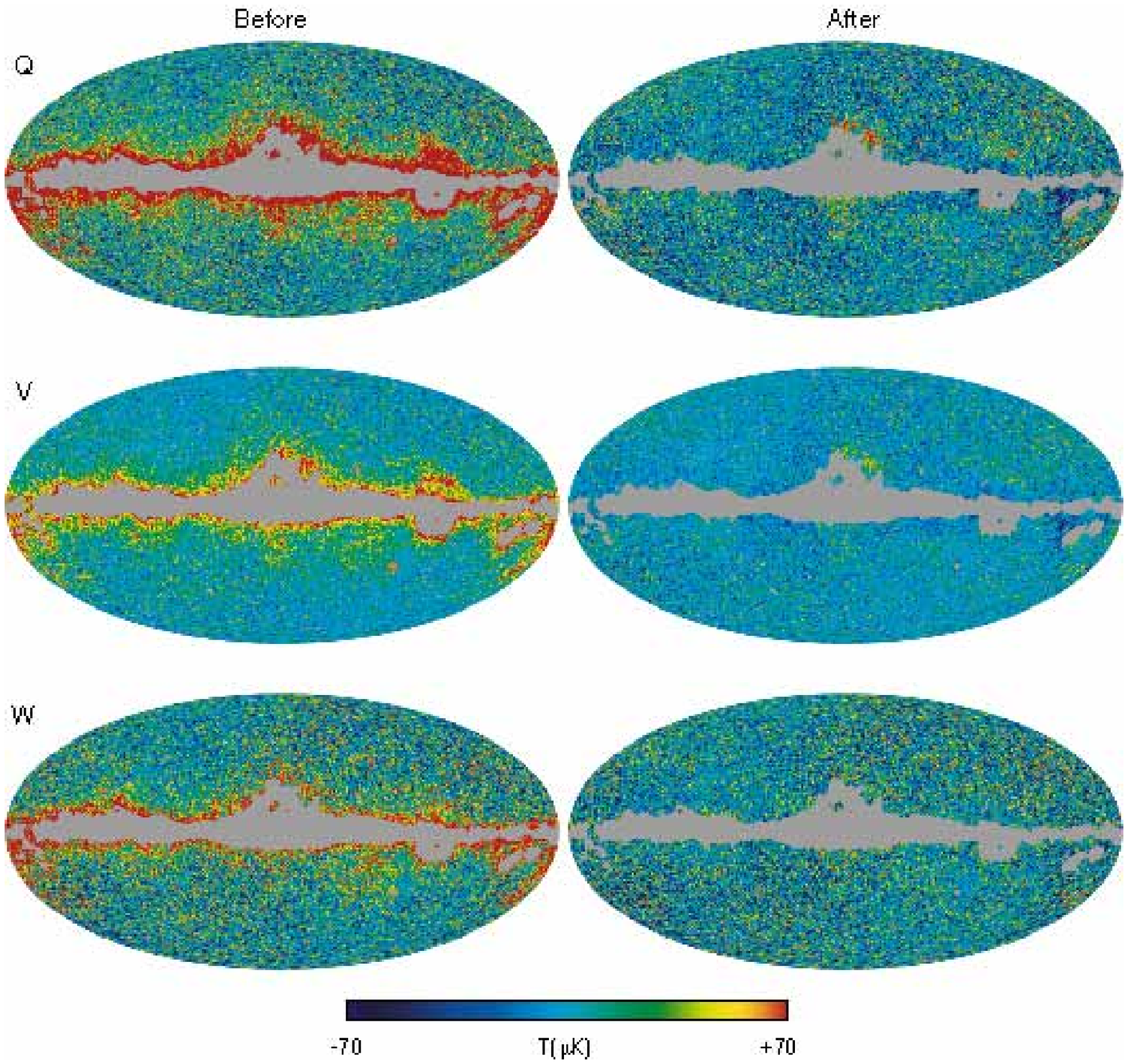}
\caption{The Q-, V-, and W-band maps, in CMB thermodynamic temperature, are shown with the Kp2 mask 
applied and a CMB estimate, from Fig \ref{bs} (a), removed.  
These maps are individually fit to template maps of synchrotron, dust, and free-free
emission to remove residual foreground contamination.  Reduction of foreground emission after 
template fitting is also shown.  The foreground reduction works well, despite the fact that 
the Haslam map is a poor tracer of synchrotron emission at microwave frequencies, because of
the similarity of the thermal dust and synchrotron morphologies.  The ``after'' template 
removal maps shown here represent the residual contamination present in the \iMAP\ CMB 
cosmological analyses.  Higher noise in the ecliptic plane is evident.
\label{templatein}} 
\end{figure}

Although this template method is not particularly physical, it does remove the
Galactic foreground outside the Kp2 cut sufficiently for CMB
analyses, as shown in Figure \ref{templatein}.  We estimate the residual 
contamination in two ways.
By examining deviations from a Gaussian distribution, we estimate that the template 
subtraction reduces the Galactic signal rms (after the sky cut)  
by 86\% in Q-band, 81\% in V-band,
and 84\% in W-band.  This agrees well with a power spectrum estimate in the range 
$2<l<100$ before and after the template subtraction.  We estimate that the 
Galactic contamination has gone from 32\% of the CMB power to 2.2\% of the power 
in Q-band, from 7.2\% to 0.8\% in V-band, and from 11\% to $<0.4$\% in W-band.
See \citet{hinshaw/etal:2003} for the power spectra before and after the template 
subtraction.

This method was adopted before the MEM
solutions were available to allow the CMB work to proceed in
parallel with the Galaxy work and thus to enable a timely release of data
products to the scientific community.  Future analyses will likely use a
modified approach, but {\it we caution that the MEM results cannot be applied
directly to CMB analyses because of their complicated noise and signal
correlations.}

We conclude that {\it template fits are a valuable method for reducing foreground 
contamination from CMB maps, but are a poor way to distinguish individual physical 
emission components due to their correlated morphological structures.}

\section{EXTRAGALACTIC SOURCES \label{pointsources}}

In addition to the Galactic foregrounds, extragalactic point sources will contaminate the
\iMAP\ anisotropy data.  Estimates of the level of 
point source contamination expected at the \iMAP\ frequencies have been made based on 
extrapolations from measured counts at higher and lower frequencies 
\citep{park/park/ratra:2002, sokasian/gawiser/smoot:2001, refregier/spergel/herbig:2000, 
toffolatti/etal:1998}.  Direct 15 GHz source count measurements 
by \citet{taylor/etal:2001} indicate that these extrapolated source counts underestimate the true
counts by a factor of two.  This is because flatter spectrum synchrotron components 
increasingly dominate over steeper spectrum components with increasing frequency, 
as in the case of Galactic emission discussed above.  
Microwave/millimeter wave observations preferentially sample flat
spectrum sources.  Techniques that remove Galactic signal 
contamination, such as the ones described above, will also generally reduce  
extragalactic contamination. For both Galactic and extragalactic contamination,
the most affected \iMAP\ pixels should be 
masked and not used for cosmological purposes. After applying 
a point source and Galactic signal minimization technique and masking the most contaminated
pixels, the residual contribution must be accounted for as a systematic error for 
CMB analyses.

We examine the point source content of the \iMAP\ maps by constructing a catalog of 
sources surveyed at 4.85 GHz using the northern celestial hemisphere 
GB6 \citep{gregory/etal:1996} catalog and the southern hemisphere PMN catalog 
\citep{griffith/etal:1994, griffith/etal:1995, wright.a/etal:1994, wright.a/etal:1996}.  
The  GB6 catalog covers $0^\circ \lt \delta \lt
+75^\circ$ to a flux limit of 18 mJy and the PMN catalog covers 
$-87^\circ \lt \delta \lt +10^\circ$ to a flux limit between 20 and 72 mJy.  Combined, these
catalogs contain 119,619 sources, with 93,799 in the region $\vert b\vert \gt 10^\circ$.  We 
examined the \iMAP\ pixel location for each of these sources as a function of their 
4.85 GHz flux density.  
We find that the detected sources are primarily flat-spectrum, with $\alpha\sim 0$.
We also considered the 15,411 IRAS objects in the infrared PSCz catalog \citep{saunders/etal:2000}, 
but found no significant signal in the \iMAP\ data.  The detected sources tend to be 
radio galaxies and quasars, not normal galaxies, so the tight infrared-radio correlation of
galaxies discussed in \S\ref{thermdust} is not relevant.

A catalog can be made of the brightest point sources in the \iMAP\ maps,
independent of their presence in external surveys.
First, the temperature maps are weighted by the square-root of
the number of observations, $N_{obs}^{1/2}$.
The maps are then filtered by $b_l/(b^2_lC^{\rm cmb}_l + C^{\rm noise})$, 
\citep{tegmark/deoliveira-costa:1998, refregier/spergel/herbig:2000}, 
where $b_l$ is the transfer function  
of the \iMAP\ beam pattern \citep{page/etal:2003b}, $C^{\rm cmb}_l$ is
the CMB angular power spectrum, and $C^{\rm noise}$ is the noise power.
Peaks that are $\gt 5\sigma$ in the filtered maps are fit to
a Gaussian profile plus a baseline plane.
This procedure results in 208 point source candidates,
listed in Table~\ref{tbl:sources}.
Once a source is identified with a
$\gt 5\sigma$ detection in any band, then flux densities are listed
for other bands if they are $\gt 2\sigma$.

We cross-correlated this catalog with the GB6, PMN, and K\"uhr catalogs,
identifying sources if they are separated by less than $11\arcmin$
(the position uncertainty is $4\arcmin$).
When more than one source lay within the cutoff radius
the brightest one was chosen, and the source flagged.
Of the 208 sources in the catalog, 203 sources have counterparts,
20 having more than one.
Since the five without counterparts are all near the detection threshold,
and simulations suggest that we should expect $5\pm4$ false detections,
these are likely spurious.
Thus, there is no evidence for a population of 
bright microwave sources without known radio counterparts.

\begin{figure}
\figurenum{12}
\epsscale{1.}
\plotone{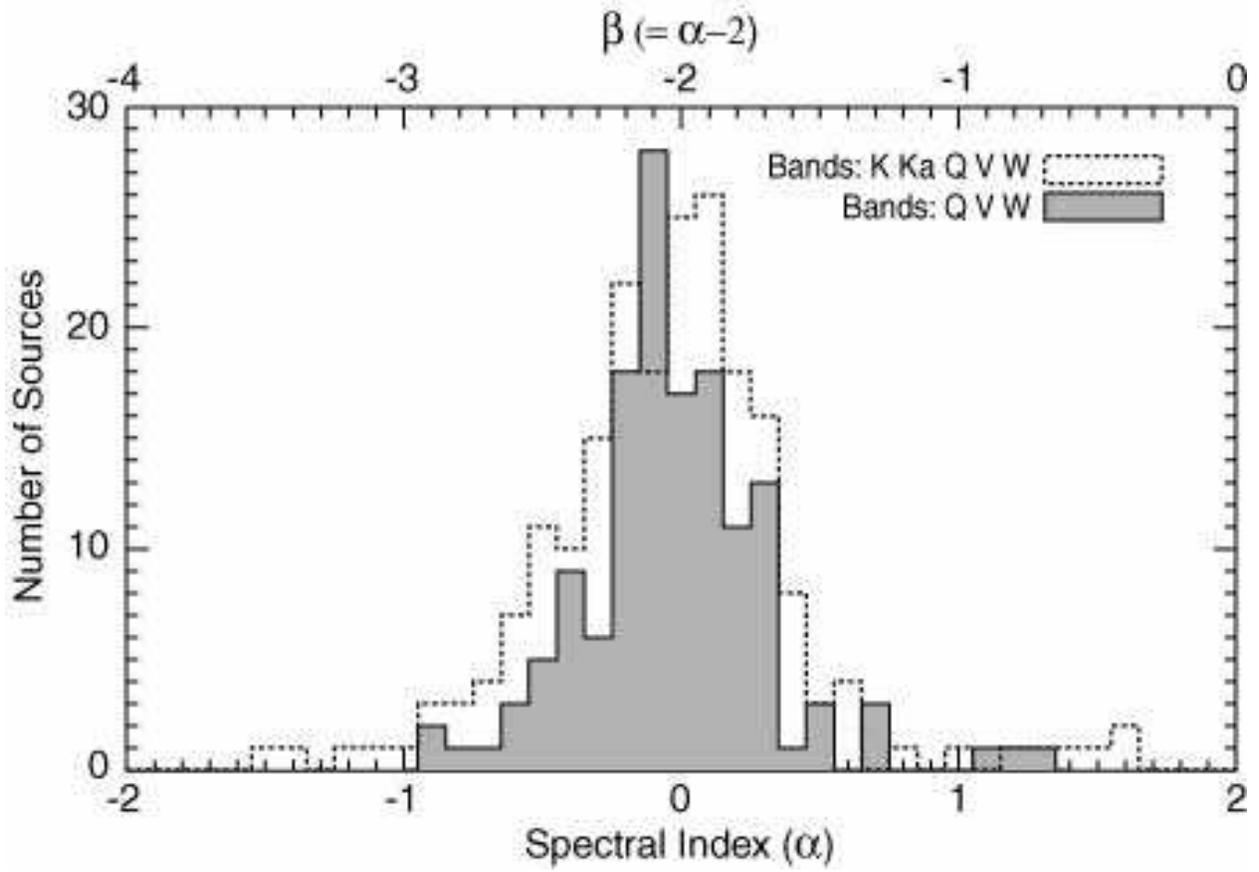}
\caption{A spectral index histogram of the 208 point sources detected by \iMAP.  
The mean spectral index is $\alpha=0$ ($\beta=-2$).
\label{fig_spectra}} 
\end{figure}

\begin{figure}
\figurenum{13}
\epsscale{1}
\plotone{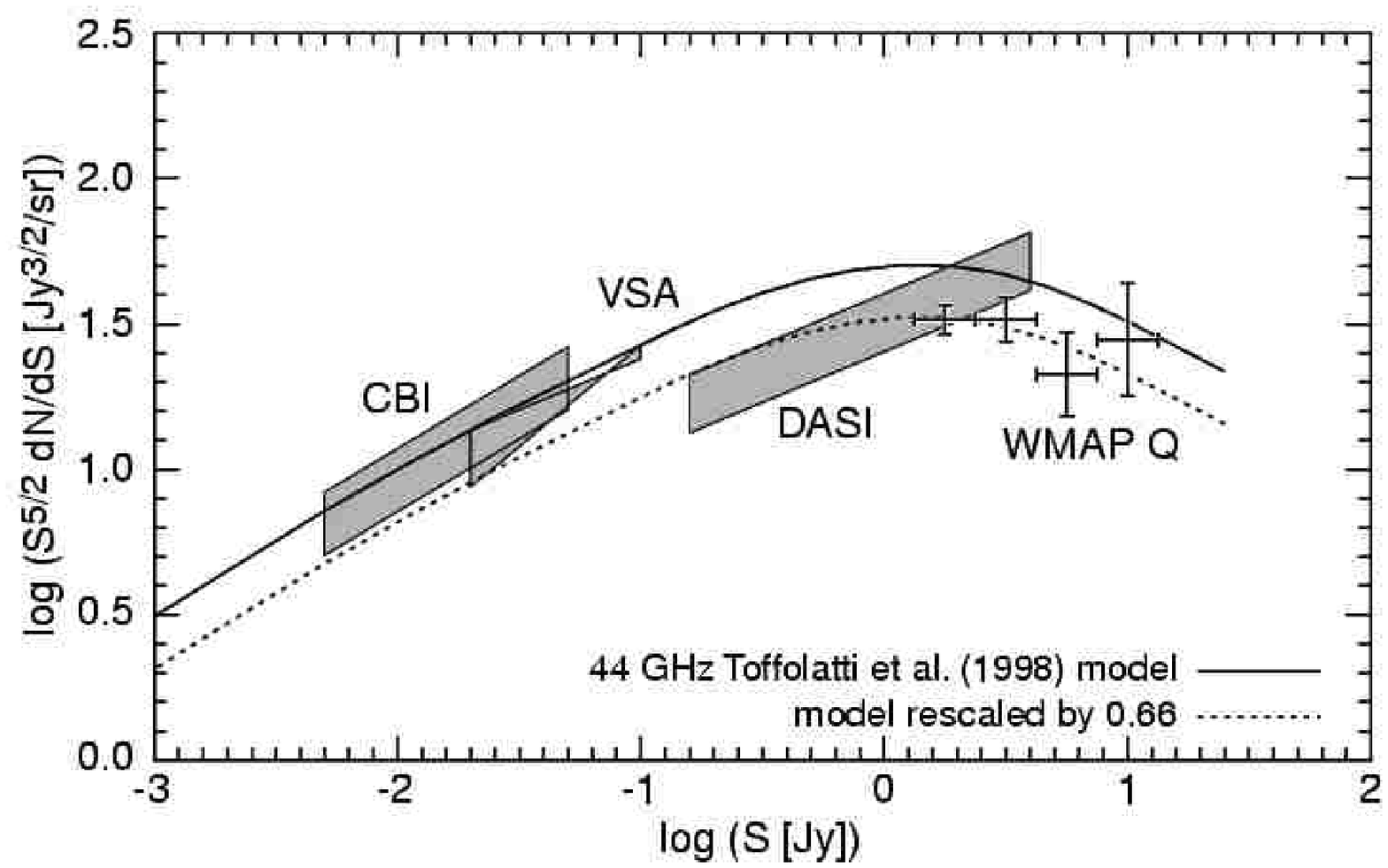}
\caption{Point source counts derived from the \iMAP\ Q-band
source catalog compared with other independent measurements and the 
\citet{toffolatti/etal:1998} 44~GHz model.
The VSA measurement of $dN/dS$ was made at 34~GHz \citep{grainge/etal:2003}.
whereas CBI \citep{mason/etal:2002}
and DASI \citep{kovac/etal:2002}
measurements were made at 31~GHz.
The CBI and DASI boxes indicate the $1\sigma$ normalization uncertainty bounds,
ignoring the slope uncertainty.
For the VSA, we show the region bounded by their quoted upper and lower
limits for $dN/dS$.
The rescaling of the Toffolatti model was found by fitting to the \iMAP\  
data alone, and not to the other experiments.
\label{fig_dnds}}
\end{figure}

The sources observed at \iMAP\ frequencies are predominately flat-spectrum
(Figure \ref{fig_spectra}), with an average spectral index of $\alpha\sim 0$.
Modeling the source count distribution $dN/dS$ as a power law
$\kappa(S/{\rm Jy})^\beta$, we obtain the values presented in
Table~\ref{tbl:src:dnds}.
We find the slopes to be nearly Euclidean, consistent with the
\citet{toffolatti/etal:1998} model (hereafter referred to as T98),
which predicts a steepening in the spectrum at fluxes greater than 1~Jy
(Figure~\ref{fig_dnds}).
T98 appears to overpredict the number of point sources;
to fit the Q-band data their model should be rescaled by $0.66\pm0.06$.
Independent empirical estimates of the source count at 90~GHz
are a factor of two below the T98 prediction
\citep{sokasian/gawiser/smoot:2001,holdaway/owen/rupen:1994}.

Given a model for the source count distribution, we can predict the level
of residual contamination due to unresolved point sources.
The contribution to the anisotropy for Poisson-distributed sources
is an effective white-noise term
\begin{equation}
C^{\rm src} = g(\nu)^2 \int_0^{S_c}{dS\,{{dN}\over{dS}}S^2}
\quad \mu{\rm K}^2 {\rm sr}
\label{eqn:psc:conflimit}
\end{equation}
where $S_c$ is the flux limit and 
$g(\nu) = \left( c^2 / 2 k \nu^2 \right) \left({{(e^x - 1)^2}/{x^2 e^x}}\right)$
converts flux density to thermodynamic temperature (with $x$ defined in 
\S\ref{intro}).
Note that point sources are constant in $C^{\rm src}$ and therefore 
have an $l^2$ dependence in $l(l+1)C_l/2\pi$.
As the point source mask is an amalgam of incomplete catalogs
it's difficult to estimate $S_c$.
Primarily constructed from a selection of sources greater than 1~Jy at 5~GHz
and 0.5~Jy at 22~GHz, a ``split-the-difference'' estimate of 0.75~Jy
seems reasonable.
This value is also suggested by measurements of the  bispectrum
\citep{komatsu/etal:2003}.
Integrating the rescaled T98 model with $S_c=0.75$ gives us
$C^{\rm src}=35.4\pm3.2\ g(\nu)^2 \;{\rm Jy}^2{\rm sr}^{-1}$,
or $(15.0\pm1.4)\times10^{-3}\mu{\rm K}^2{\rm sr}$ for Q-band.
Extrapolating, this implies nearly negligible contamination levels
of $3$ and $1\times10^{-3}\mu{\rm K}^2{\rm sr}$ for V- and W-band.

\section{SUNYAEV-ZELDOVICH (SZ) EFFECT \label{sz}}

Hot gas in clusters of galaxies will contaminate the maps by shifting the spectrum of the 
primary anisotropy to create a Sunyaev-Zeldovich decrement in the \iMAP\ frequency 
bands. This is, however, a small effect for \iMAP.  

The brightest SZ source is the Coma cluster, which is included in the point source mask 
described in \S\ref{mask}.  The Coma decrement observed by \iMAP\ is $-0.34\pm 0.18$ mK in 
W-band and $-0.24\pm 0.18$ mK in V-band.  

More generally, the XBACs catalog of 242 Abell-type clusters
\citep{ebeling/etal:1996} is correlated with the \iMAP\ W-band map.
The XBACs clusters are treated as point sources for \iMAP\ and the conversion
to units of flux density is based on \citet{refregier/spergel/herbig:2000},
using the relation
\begin{equation}
S_{94} = 11.44 \left({{300\,{\rm Mpc}}\over{D(z)}}\right)^2
        \left({{f_{\rm gas}}\over{0.11}}\right)
        \left({{kT_e}\over{1\,{\rm keV}}}\right)^{5/2}
        \quad[{\rm mJy}]
\end{equation}
where $D(z)$ is the angular diameter distance to the cluster,
$f_{\rm gas}\equiv M_{\rm gas}/M$ is the gas mass fraction, and
$T_e$ is the electron temperature.
The overall normalization of this relation is uncertain due to ignorance
of the correct gas fraction and cluster virialization state.
We fix $f_{\rm gas} = 0.11$.
Extended clusters (more than one-third the extent of Coma) are omitted;
the remaining fluxes are all $<1$ Jy.
A template map is constructed by convolving the clusters with the \iMAP\ beam
\citep{page/etal:2003b}, and then the template is fit to the \iMAP\ W-band map.
We find a template normalization of $-0.36\pm 0.14$.
Since the fluxes used to construct the template were positive, the negative
scaling is consistent with observing the SZ effect at 2.5$\sigma$ 
using a matched filter (the X-ray catalog).

CMB photons that travel to us through the plane of our own Galaxy undergo an SZ 
distortion of $y\approx k n_e T_e \sigma_T L / m_e c^2$, where $\sigma_T$ is the Thomson 
scattering cross-section and $L$ is the effective (electron pressure weighted) length 
of the path through our Galaxy.  Taking $n_e T_e = 10^3$ K cm$^{-3}$ and $L=50$ kpc we get  
$y\approx 2\times 10^{-8}$.  Therefore our Galaxy does not significantly affect the 
CMB photons, even in the worst case.  More generally, the SZ effect, as a contaminating 
foreground, can safely be ignored for \iMAP\ data analyses.

\clearpage
\begin{deluxetable}{ccccccccccc}
\tablecaption{\label{tbl:sources} MAP Source Catalog}
\tabletypesize{\footnotesize}
\tablewidth{0pt}
\tablehead{
&\colhead{$l$ [$\arcdeg$]}
&\colhead{$b$ [$\arcdeg$]}
&\colhead{K [Jy]}
&\colhead{Ka [Jy]}
&\colhead{Q [Jy]}
&\colhead{V [Jy]}
&\colhead{W [Jy]}
&\colhead{$\alpha$}
&\colhead{5 GHz ID}
&
}
\startdata
001
&$0.59$
&$-42.88$
&$1.5\pm0.1$
&$1.8\pm0.2$
&$1.4\pm0.3$
&$1.0\pm0.3$
&\nodata
&$-0.0\pm0.5$
&PMN J2109-4110
&
\\
002
&$1.31$
&\phs$45.99$
&$1.8\pm0.1$
&$2.2\pm0.2$
&$1.6\pm0.2$
&$1.1\pm0.4$
&\nodata
&$-0.0\pm0.4$
&GB6 J1516+0015
&
\\
003
&$1.58$
&$-28.97$
&$3.2\pm0.1$
&$3.7\pm0.2$
&$2.9\pm0.2$
&$2.6\pm0.4$
&$2.8\pm0.7$
&$-0.1\pm0.2$
&PMN J1957-3845
&
\\
004
&$8.95$
&\phs$73.13$
&$1.3\pm0.1$
&$1.5\pm0.2$
&$1.3\pm0.2$
&\nodata
&\nodata
&$0.0\pm0.7$
&GB6 J1357+1919
&
\\
005
&$10.85$
&\phs$40.92$
&$1.9\pm0.1$
&$1.9\pm0.2$
&$1.7\pm0.3$
&$1.6\pm0.6$
&\nodata
&$-0.1\pm0.5$
&GB6 J1549+0237
&
\\
006
&$11.38$
&\phs$54.54$
&$2.1\pm0.1$
&$2.1\pm0.2$
&$2.2\pm0.2$
&\nodata
&\nodata
&$0.1\pm0.4$
&GB6 J1504+1029
&
\\
007
&$14.22$
&\phs$42.20$
&$2.3\pm0.1$
&$2.2\pm0.2$
&$1.8\pm0.2$
&$1.9\pm0.5$
&\nodata
&$-0.2\pm0.4$
&GB6 J1550+0527
&
\\
008
&$17.14$
&$-16.25$
&$2.4\pm0.1$
&$2.5\pm0.2$
&$2.3\pm0.2$
&\nodata
&\nodata
&$0.0\pm0.4$
&PMN J1923-2104
&
\\
009
&$23.01$
&\phs$40.76$
&$3.2\pm0.1$
&$3.3\pm0.2$
&$2.7\pm0.2$
&$3.1\pm0.5$
&$3.1\pm0.8$
&$-0.1\pm0.2$
&GB6 J1608+1029
&
\\
010
&$23.08$
&\phs$28.93$
&$1.1\pm0.1$
&$0.7\pm0.2$
&\nodata
&\nodata
&\nodata
&$-1.4\pm2$
&GB6 J1651+0459
&
\\
011
&$24.00$
&$-23.13$
&$1.5\pm0.1$
&$1.2\pm0.3$
&$1.3\pm0.2$
&$1.7\pm0.3$
&\nodata
&$0.0\pm0.5$
&PMN J2000-1748
&
\\
012
&$24.41$
&$-64.92$
&$9.6\pm0.1$
&$9.8\pm0.2$
&$9.8\pm0.2$
&$8.9\pm0.5$
&$6.3\pm0.8$
&$-0.0\pm0.08$
&PMN J2258-2758
&
\\
013
&$26.71$
&\phs$28.71$
&$1.3\pm0.1$
&$1.3\pm0.2$
&$1.2\pm0.2$
&\nodata
&\nodata
&$-0.1\pm0.7$
&GB6 J1658+0741
&
\\
014
&$27.04$
&$-24.61$
&\nodata
&$0.7\pm0.2$
&\nodata
&$1.8\pm0.6$
&\nodata
&$1.6\pm2$
&PMN J2011-1546
&
\\
015
&$32.73$
&$-29.65$
&$0.7\pm0.2$
&\nodata
&$1.0\pm0.2$
&\nodata
&\nodata
&$0.5\pm1$
&\nodata
&
\\
016
&$36.61$
&$-51.20$
&$1.9\pm0.1$
&$1.2\pm0.2$
&$1.9\pm0.3$
&$1.6\pm0.6$
&\nodata
&$-0.3\pm0.5$
&PMN J2206-1835
&
\\
017
&$40.54$
&$-40.95$
&$3.2\pm0.1$
&$2.7\pm0.2$
&$2.9\pm0.2$
&$2.0\pm0.4$
&\nodata
&$-0.3\pm0.3$
&PMN J2131-1207
&
\\
018
&$40.61$
&$-48.05$
&$2.1\pm0.1$
&$2.3\pm0.2$
&$2.1\pm0.3$
&$2.1\pm0.6$
&\nodata
&$0.0\pm0.4$
&PMN J2158-1501
&
\\
019
&$41.54$
&$-62.81$
&$1.2\pm0.2$
&$1.3\pm0.2$
&\nodata
&$1.9\pm0.8$
&\nodata
&$0.3\pm0.8$
&PMN J2256-2011
&
\\
020
&$52.32$
&$-36.49$
&$1.7\pm0.1$
&$1.7\pm0.2$
&$1.6\pm0.3$
&$1.4\pm0.5$
&$2.0\pm0.6$
&$-0.0\pm0.4$
&PMN J2134-0153
&
\\
021
&$53.78$
&$-57.10$
&$1.6\pm0.1$
&$1.4\pm0.3$
&$1.5\pm0.4$
&\nodata
&\nodata
&$-0.2\pm0.8$
&PMN J2246-1206
&
\\
022
&$54.15$
&\phs$24.49$
&$2.1\pm0.09$
&$2.2\pm0.2$
&$2.7\pm0.2$
&$3.2\pm0.4$
&$2.2\pm0.7$
&$0.3\pm0.2$
&GB6 J1753+2847
&
\\
023
&$55.14$
&\phs$46.37$
&$3.5\pm0.1$
&$2.8\pm0.2$
&$2.9\pm0.2$
&$2.0\pm0.4$
&$1.4\pm0.5$
&$-0.5\pm0.2$
&GB6 J1613+3412
&
\\
024
&$55.21$
&$-51.70$
&$1.5\pm0.1$
&$1.9\pm0.2$
&$2.2\pm0.3$
&$2.2\pm0.3$
&\nodata
&$0.4\pm0.4$
&PMN J2229-0832
&
\\
025
&$55.48$
&$-35.57$
&$4.2\pm0.1$
&$3.6\pm0.2$
&$3.2\pm0.3$
&$2.3\pm0.5$
&\nodata
&$-0.5\pm0.3$
&GB6 J2136+0041
&
\\
026
&$56.47$
&\phs$80.64$
&$2.1\pm0.1$
&$2.1\pm0.2$
&$1.3\pm0.2$
&\nodata
&\nodata
&$-0.5\pm0.5$
&GB6 J1331+3030
&
\\
027
&$58.00$
&$-30.11$
&$2.1\pm0.1$
&$1.7\pm0.2$
&$2.0\pm0.3$
&\nodata
&$2.1\pm0.8$
&$-0.1\pm0.4$
&GB6 J2123+0535
&
\\
028
&$58.57$
&\phs$12.64$
&$1.4\pm0.08$
&$1.1\pm0.1$
&$0.9\pm0.1$
&\nodata
&\nodata
&$-0.6\pm0.5$
&GB6 J1850+2825
&
\\
029
&$58.97$
&$-48.84$
&$6.4\pm0.1$
&$6.1\pm0.2$
&$5.7\pm0.3$
&$4.8\pm0.5$
&$7.1\pm3$
&$-0.2\pm0.1$
&PMN J2225-0457
&
\\
030
&$59.03$
&$-46.63$
&$2.5\pm0.1$
&$1.8\pm0.2$
&$1.8\pm0.3$
&\nodata
&\nodata
&$-0.7\pm0.6$
&PMN J2218-0335
&
\\
031
&$59.31$
&$-11.63$
&$1.0\pm0.1$
&$1.6\pm0.2$
&\nodata
&\nodata
&\nodata
&$1.2\pm0.9$
&GB6 J2024+1718
&
\\
032
&$59.85$
&$-68.32$
&$1.5\pm0.1$
&\nodata
&$1.5\pm0.3$
&\nodata
&$1.9\pm0.6$
&$0.2\pm0.5$
&PMN J2331-1556
&
\\
033
&$61.08$
&\phs$42.33$
&$4.6\pm0.1$
&$5.6\pm0.2$
&$6.0\pm0.2$
&$5.9\pm0.3$
&$4.2\pm0.7$
&$0.3\pm0.1$
&GB6 J1635+3808
&
\\
034
&$62.93$
&\phs$11.65$
&$1.3\pm0.09$
&$1.3\pm0.2$
&$0.8\pm0.1$
&$0.9\pm0.4$
&\nodata
&$-0.5\pm0.6$
&GB6 J1902+3159
&
\\
035
&$63.45$
&\phs$40.96$
&$8.0\pm0.1$
&$7.4\pm0.2$
&$6.8\pm0.2$
&$5.8\pm0.4$
&$5.7\pm0.7$
&$-0.3\pm0.09$
&GB6 J1642+3948
&
\\
036
&$63.45$
&\phs$38.78$
&$1.2\pm0.09$
&$1.4\pm0.2$
&$0.5\pm0.2$
&\nodata
&\nodata
&$-0.1\pm0.8$
&GB6 J1653+3945
&*
\\
037
&$63.66$
&$-34.07$
&$8.6\pm0.1$
&$8.2\pm0.2$
&$8.1\pm0.3$
&$6.1\pm0.4$
&\nodata
&$-0.2\pm0.1$
&GB6 J2148+0657
&
\\
038
&$63.91$
&\phs$31.02$
&$1.1\pm0.1$
&\nodata
&$1.4\pm0.2$
&$1.1\pm0.4$
&\nodata
&$0.2\pm0.5$
&GB6 J1734+3857
&
\\
039
&$65.57$
&$-71.92$
&$2.0\pm0.1$
&$2.1\pm0.2$
&$2.3\pm0.2$
&$2.1\pm0.7$
&$2.3\pm0.7$
&$0.1\pm0.3$
&PMN J2348-1631
&
\\
040
&$65.30$
&\phs$80.35$
&$1.1\pm0.1$
&$1.4\pm0.3$
&$1.4\pm0.6$
&\nodata
&\nodata
&$0.5\pm1$
&GB6 J1329+3154
&
\\
041
&$68.60$
&$-27.57$
&$2.6\pm0.1$
&$2.4\pm0.2$
&$1.9\pm0.2$
&$1.7\pm0.3$
&$1.5\pm0.8$
&$-0.4\pm0.3$
&GB6 J2139+1423
&
\\
042
&$69.83$
&\phs$68.37$
&\nodata
&$1.2\pm0.1$
&$1.3\pm0.2$
&$1.6\pm0.3$
&$1.5\pm0.5$
&$0.3\pm0.7$
&GB6 J1419+3822
&
\\
043
&$71.44$
&\phs$33.30$
&$1.7\pm0.1$
&$2.0\pm0.2$
&$1.4\pm0.2$
&$2.0\pm0.3$
&$1.6\pm0.5$
&$0.0\pm0.3$
&GB6 J1727+4530
&
\\
044
&$72.02$
&$-26.11$
&$1.3\pm0.1$
&$1.4\pm0.2$
&\nodata
&$2.0\pm0.8$
&$1.2\pm0.6$
&$0.1\pm0.6$
&GB6 J2143+1743
&
\\
045
&$75.60$
&$-29.66$
&$1.2\pm0.1$
&$1.3\pm0.2$
&$1.4\pm0.2$
&$0.9\pm0.3$
&\nodata
&$0.0\pm0.5$
&GB6 J2203+1725
&
\\
046
&$77.25$
&\phs$23.49$
&$2.6\pm0.1$
&$2.6\pm0.2$
&$2.3\pm0.2$
&$1.5\pm0.2$
&$1.7\pm0.4$
&$-0.3\pm0.2$
&GB6 J1829+4844
&
\\
047
&$77.47$
&$-38.57$
&$3.6\pm0.1$
&$3.5\pm0.2$
&$3.7\pm0.3$
&$3.6\pm0.4$
&$3.1\pm0.7$
&$-0.0\pm0.2$
&GB6 J2232+1143
&
\\
048
&$79.57$
&\phs$31.77$
&$1.1\pm0.1$
&$1.1\pm0.1$
&$0.4\pm0.2$
&$1.8\pm0.4$
&\nodata
&$0.2\pm0.5$
&GB6 J1740+5211
&
\\
049
&$80.36$
&$-8.35$
&\nodata
&$1.3\pm0.2$
&$0.9\pm0.2$
&$1.4\pm0.3$
&$1.7\pm0.6$
&$0.3\pm0.7$
&GB6 J2109+3532
&*
\\
050
&$82.11$
&$-26.09$
&$1.3\pm0.1$
&\nodata
&$1.7\pm0.2$
&$1.3\pm0.6$
&\nodata
&$0.3\pm0.5$
&GB6 J2212+2355
&
\\
051
&$85.34$
&\phs$11.82$
&$0.5\pm0.09$
&$1.2\pm0.2$
&$1.2\pm0.2$
&\nodata
&\nodata
&$1.5\pm0.9$
&GB6 J1955+5131
&
\\
052
&$85.67$
&\phs$83.32$
&$2.6\pm0.1$
&$2.5\pm0.2$
&$2.6\pm0.2$
&$2.2\pm0.4$
&\nodata
&$-0.1\pm0.3$
&GB6 J1310+3220
&
\\
053
&$85.73$
&\phs$26.04$
&$1.4\pm0.09$
&$1.3\pm0.2$
&$1.5\pm0.2$
&\nodata
&\nodata
&$0.1\pm0.5$
&GB6 J1824+5650
&
\\
054
&$85.96$
&$-18.76$
&$3.5\pm0.1$
&$3.3\pm0.2$
&$3.0\pm0.2$
&$2.4\pm0.4$
&$3.7\pm0.8$
&$-0.2\pm0.2$
&GB6 J2203+3145
&
\\
055
&$86.13$
&$-38.19$
&$7.5\pm0.1$
&$7.1\pm0.2$
&$6.9\pm0.2$
&$5.9\pm0.4$
&$5.9\pm0.9$
&$-0.2\pm0.1$
&GB6 J2253+1608
&
\\
056
&$86.73$
&\phs$40.33$
&\nodata
&$1.7\pm0.2$
&$1.1\pm0.1$
&$1.3\pm0.2$
&\nodata
&$-0.5\pm0.8$
&GB6 J1638+5720
&
\\
057
&$90.07$
&$-25.65$
&$1.4\pm0.1$
&$1.9\pm0.2$
&$1.3\pm0.2$
&\nodata
&\nodata
&$0.2\pm0.6$
&GB6 J2236+2828
&
\\
058
&$92.62$
&$-10.44$
&$3.1\pm0.1$
&$2.9\pm0.2$
&$3.3\pm0.2$
&\nodata
&\nodata
&$0.1\pm0.2$
&GB6 J2202+4216
&
\\
059
&$92.74$
&\phs$19.45$
&\nodata
&$1.0\pm0.2$
&$1.4\pm0.2$
&$1.2\pm0.3$
&$1.2\pm0.6$
&$0.2\pm0.8$
&GB6 J1927+6117
&
\\
060
&$93.45$
&$-66.61$
&$2.0\pm0.2$
&$2.1\pm0.3$
&$1.9\pm0.3$
&$2.0\pm0.4$
&\nodata
&$0.0\pm0.4$
&PMN J0006-0623
&*
\\
061
&$96.01$
&$-11.23$
&$2.2\pm0.5$
&$1.4\pm0.3$
&\nodata
&\nodata
&\nodata
&$-1.2\pm2$
&\nodata
&
\\
062
&$96.87$
&$-87.87$
&$1.3\pm0.1$
&$0.4\pm0.2$
&$1.3\pm0.3$
&$1.9\pm0.6$
&\nodata
&$0.2\pm0.6$
&PMN J0047-2517
&
\\
063
&$96.09$
&\phs$13.72$
&$1.7\pm0.1$
&$1.5\pm0.2$
&$1.4\pm0.2$
&$1.3\pm0.3$
&\nodata
&$-0.3\pm0.4$
&GB6 J2022+6137
&
\\
064
&$96.33$
&\phs$29.90$
&$0.6\pm0.04$
&$0.6\pm0.05$
&$0.5\pm0.1$
&$0.5\pm0.2$
&\nodata
&$-0.2\pm0.5$
&GB6 J1758+6638
&*
\\
065
&$97.50$
&\phs$25.00$
&$1.2\pm0.08$
&$1.4\pm0.1$
&$1.6\pm0.1$
&$1.1\pm0.2$
&$1.4\pm0.4$
&$0.1\pm0.3$
&GB6 J1849+6705
&*
\\
066
&$98.48$
&\phs$25.81$
&\nodata
&$1.1\pm0.1$
&$1.1\pm0.1$
&\nodata
&\nodata
&$0.3\pm2$
&GB6 J1842+6809
&
\\
067
&$100.13$
&\phs$29.17$
&$1.3\pm0.07$
&$1.1\pm0.1$
&$1.2\pm0.1$
&$1.5\pm0.3$
&\nodata
&$-0.0\pm0.3$
&GB6 J1806+6949
&
\\
068
&$100.54$
&\phs$30.69$
&$0.6\pm0.09$
&$2.0\pm0.8$
&$1.0\pm0.1$
&$1.5\pm0.3$
&\nodata
&$0.8\pm0.5$
&GB6 J1748+7005
&
\\
069
&$100.59$
&\phs$36.63$
&$1.2\pm0.1$
&$0.9\pm0.2$
&$0.9\pm0.2$
&\nodata
&\nodata
&$-0.5\pm0.8$
&GB6 J1642+6856
&*
\\
070
&$105.62$
&\phs$23.55$
&$3.7\pm0.09$
&$3.8\pm0.1$
&$3.2\pm0.1$
&$3.5\pm0.3$
&\nodata
&$-0.1\pm0.1$
&GB6 J1927+7357
&
\\
071
&$110.04$
&\phs$42.11$
&$1.6\pm0.1$
&$1.4\pm0.2$
&$1.5\pm0.2$
&$1.2\pm0.5$
&\nodata
&$-0.2\pm0.5$
&GB6 J1459+7140
&
\\
072
&$110.04$
&\phs$29.09$
&$2.3\pm0.1$
&$1.8\pm0.1$
&$2.3\pm0.2$
&$1.5\pm0.3$
&\nodata
&$-0.2\pm0.3$
&1Jy 1803+78
&
\\
073
&$111.41$
&\phs$27.14$
&$1.3\pm0.09$
&$1.0\pm0.2$
&\nodata
&\nodata
&\nodata
&$-0.7\pm1$
&1Jy 1845+79
&
\\
074
&$112.64$
&$-15.89$
&$1.8\pm0.1$
&\nodata
&$1.7\pm0.3$
&$1.8\pm0.6$
&\nodata
&$-0.1\pm0.5$
&GB6 J2354+4553
&
\\
075
&$113.92$
&$-12.04$
&$1.0\pm0.08$
&$0.9\pm0.1$
&\nodata
&\nodata
&\nodata
&$-0.2\pm0.9$
&\nodata
&
\\
076
&$115.66$
&\phs$31.23$
&$1.2\pm0.09$
&$1.5\pm0.1$
&$1.0\pm0.1$
&$1.2\pm0.3$
&\nodata
&$0.1\pm0.4$
&\nodata
&
\\
077
&$122.77$
&$-72.30$
&$1.3\pm0.1$
&$1.4\pm0.2$
&\nodata
&\nodata
&\nodata
&$0.1\pm1$
&PMN J0050-0928
&
\\
078
&$125.71$
&\phs$35.79$
&\nodata
&\nodata
&$0.6\pm0.2$
&$1.5\pm0.6$
&\nodata
&$2.5\pm3$
&1Jy 1150+81
&
\\
079
&$129.26$
&$-49.33$
&$1.4\pm0.1$
&$1.6\pm0.3$
&\nodata
&\nodata
&\nodata
&$0.4\pm1$
&GB6 J0108+1319
&
\\
080
&$130.79$
&$-14.29$
&$4.5\pm0.1$
&$4.7\pm0.2$
&$4.2\pm0.2$
&$3.3\pm0.4$
&$2.9\pm0.7$
&$-0.2\pm0.2$
&GB6 J0136+4751
&
\\
081
&$131.83$
&$-60.97$
&$2.1\pm0.1$
&$2.3\pm0.2$
&$2.3\pm0.3$
&$2.5\pm0.4$
&\nodata
&$0.2\pm0.3$
&GB6 J0108+0135
&*
\\
082
&$133.47$
&\phs$18.41$
&$1.0\pm0.1$
&$1.4\pm0.3$
&$0.6\pm0.2$
&$0.9\pm0.4$
&\nodata
&$-0.3\pm0.8$
&1Jy 0403+76
&
\\
083
&$135.56$
&\phs$42.27$
&$1.1\pm0.1$
&$1.5\pm0.3$
&$1.1\pm0.3$
&$1.4\pm0.5$
&\nodata
&$0.2\pm0.7$
&GB6 J1048+7143
&
\\
084
&$140.19$
&$-16.72$
&$2.1\pm0.1$
&\nodata
&$1.3\pm0.3$
&$1.3\pm0.4$
&$1.8\pm0.7$
&$-0.4\pm0.4$
&GB6 J0223+4259
&*
\\
085
&$140.50$
&$-28.14$
&$1.5\pm0.2$
&$1.8\pm0.3$
&$1.3\pm0.4$
&\nodata
&\nodata
&$0.1\pm1$
&GB6 J0205+3212
&
\\
086
&$141.17$
&$-61.85$
&$1.3\pm0.1$
&$1.7\pm0.2$
&$1.4\pm0.3$
&$1.8\pm0.4$
&\nodata
&$0.3\pm0.5$
&PMN J0125-0005
&*
\\
087
&$141.27$
&\phs$25.11$
&$0.8\pm0.1$
&$0.8\pm0.2$
&$1.1\pm0.2$
&\nodata
&\nodata
&$0.4\pm0.8$
&GB6 J0639+7324
&
\\
088
&$141.53$
&\phs$40.59$
&$1.4\pm0.1$
&$1.3\pm0.2$
&$1.3\pm0.2$
&\nodata
&\nodata
&$-0.2\pm0.7$
&GB6 J0955+6940
&
\\
089
&$143.48$
&\phs$34.43$
&$1.5\pm0.1$
&\nodata
&$1.2\pm0.2$
&$1.4\pm0.4$
&\nodata
&$-0.2\pm0.5$
&GB6 J0841+7053
&
\\
090
&$145.59$
&\phs$64.97$
&$1.8\pm0.09$
&\nodata
&$1.8\pm0.2$
&$1.6\pm0.4$
&$1.3\pm0.6$
&$-0.1\pm0.4$
&GB6 J1153+4931
&*
\\
091
&$146.82$
&\phs$20.78$
&$1.2\pm0.08$
&$1.2\pm0.1$
&$1.3\pm0.4$
&$1.3\pm0.4$
&\nodata
&$0.1\pm0.5$
&GB6 J0607+6720
&
\\
092
&$147.92$
&$-44.04$
&$1.7\pm0.2$
&$1.6\pm0.3$
&$1.8\pm0.2$
&\nodata
&\nodata
&$0.0\pm0.6$
&GB6 J0204+1514
&
\\
093
&$149.49$
&$-28.52$
&$3.9\pm0.1$
&$4.0\pm0.2$
&$3.9\pm0.3$
&$1.8\pm0.4$
&\nodata
&$-0.2\pm0.3$
&GB6 J0237+2848
&
\\
094
&$150.56$
&$-13.26$
&$11.1\pm0.1$
&$8.7\pm0.2$
&$6.9\pm0.3$
&$4.7\pm0.4$
&\nodata
&$-0.8\pm0.1$
&GB6 J0319+4130
&
\\
095
&$161.69$
&\phs$10.31$
&$1.6\pm0.2$
&$1.4\pm0.3$
&\nodata
&\nodata
&\nodata
&$-0.5\pm1$
&GB6 J0542+4951
&
\\
096
&$162.36$
&$-54.39$
&$1.6\pm0.1$
&$3.4\pm2$
&$1.6\pm0.2$
&\nodata
&\nodata
&$0.1\pm0.5$
&GB6 J0217+0144
&
\\
097
&$167.92$
&$-76.02$
&$0.6\pm0.1$
&$1.6\pm0.2$
&$1.9\pm0.2$
&$1.6\pm0.6$
&\nodata
&$1.4\pm0.7$
&PMN J0132-1654
&
\\
098
&$170.20$
&\phs$50.71$
&$1.7\pm0.1$
&$1.7\pm0.2$
&$1.2\pm0.2$
&\nodata
&\nodata
&$-0.4\pm0.6$
&GB6 J0958+4725
&
\\
099
&$171.12$
&\phs$17.94$
&$3.2\pm0.1$
&$2.4\pm0.2$
&$2.5\pm0.3$
&$2.3\pm0.5$
&$1.7\pm0.8$
&$-0.5\pm0.3$
&GB6 J0646+4451
&
\\
100
&$171.73$
&\phs$7.30$
&$1.2\pm0.2$
&$2.2\pm0.2$
&\nodata
&\nodata
&\nodata
&$1.6\pm1$
&GB6 J0555+3948
&
\\
101
&$174.61$
&\phs$69.73$
&$1.3\pm0.1$
&$1.0\pm0.2$
&$1.6\pm0.4$
&$2.8\pm0.7$
&\nodata
&$0.5\pm0.5$
&GB6 J1130+3815
&*
\\
102
&$174.96$
&$-44.53$
&$1.5\pm0.2$
&$1.2\pm0.2$
&$1.2\pm0.2$
&$1.0\pm0.4$
&\nodata
&$-0.4\pm0.6$
&GB6 J0308+0406
&
\\
103
&$177.42$
&\phs$58.33$
&$1.5\pm0.1$
&$1.7\pm0.2$
&$1.5\pm0.3$
&\nodata
&$2.6\pm0.9$
&$0.2\pm0.4$
&GB6 J1033+4115
&
\\
104
&$181.05$
&\phs$50.30$
&$1.5\pm0.1$
&$1.8\pm0.2$
&$1.8\pm0.2$
&$2.0\pm0.4$
&\nodata
&$0.3\pm0.4$
&GB6 J0948+4039
&
\\
105
&$183.72$
&\phs$46.16$
&$6.8\pm0.1$
&$5.8\pm0.2$
&$5.5\pm0.2$
&$5.1\pm0.5$
&$3.1\pm0.8$
&$-0.4\pm0.1$
&GB6 J0927+3902
&
\\
106
&$187.97$
&$-42.46$
&$2.8\pm0.2$
&$2.8\pm0.3$
&$3.1\pm0.3$
&$3.0\pm0.7$
&$3.2\pm0.8$
&$0.1\pm0.3$
&PMN J0339-0146
&
\\
107
&$188.62$
&\phs$23.62$
&$1.6\pm0.1$
&$0.8\pm0.2$
&\nodata
&$2.3\pm0.9$
&$2.2\pm1$
&$0.0\pm0.5$
&GB6 J0741+3112
&
\\
108
&$190.39$
&$-27.38$
&$2.2\pm0.1$
&$2.4\pm0.2$
&$1.9\pm0.3$
&$1.8\pm0.6$
&$3.0\pm1$
&$0.0\pm0.4$
&GB6 J0433+0521
&
\\
109
&$193.60$
&$-31.74$
&$1.6\pm0.2$
&$2.1\pm0.2$
&$1.9\pm0.2$
&\nodata
&\nodata
&$0.4\pm0.6$
&GB6 J0424+0036
&
\\
110
&$195.30$
&$-33.13$
&$7.4\pm0.1$
&$7.3\pm0.2$
&$7.0\pm0.3$
&$6.8\pm0.5$
&$3.9\pm1$
&$-0.1\pm0.1$
&PMN J0423-0120
&
\\
111
&$199.35$
&\phs$78.38$
&$3.5\pm0.1$
&$3.6\pm0.2$
&$3.4\pm0.2$
&$3.0\pm0.4$
&$2.1\pm0.6$
&$-0.1\pm0.2$
&GB6 J1159+2914
&
\\
112
&$200.07$
&\phs$31.86$
&$1.7\pm0.2$
&$1.6\pm0.2$
&$2.4\pm0.4$
&$3.1\pm0.7$
&\nodata
&$0.5\pm0.5$
&GB6 J0830+2410
&
\\
113
&$201.73$
&\phs$18.16$
&$1.5\pm0.1$
&\nodata
&$1.5\pm0.3$
&\nodata
&\nodata
&$0.1\pm0.7$
&GB6 J0738+1742
&
\\
114
&$205.62$
&$-42.60$
&$4.8\pm1$
&$1.6\pm0.2$
&$1.6\pm0.2$
&$1.3\pm0.4$
&\nodata
&$-1.0\pm0.8$
&PMN J0405-1308
&
\\
115
&$206.82$
&\phs$35.80$
&$2.6\pm0.2$
&$3.5\pm0.3$
&$2.9\pm0.3$
&$2.5\pm0.5$
&$2.5\pm0.8$
&$0.1\pm0.3$
&GB6 J0854+2006
&
\\
116
&$207.39$
&$-23.18$
&\nodata
&\nodata
&$1.7\pm0.2$
&\nodata
&$1.8\pm0.8$
&$0.0\pm1$
&\nodata
&
\\
117
&$208.20$
&\phs$18.76$
&$2.3\pm0.2$
&$2.0\pm0.2$
&$2.1\pm0.3$
&$2.1\pm0.5$
&\nodata
&$-0.2\pm0.4$
&GB6 J0750+1231
&
\\
118
&$209.74$
&\phs$16.58$
&$1.4\pm0.2$
&\nodata
&$0.9\pm0.2$
&\nodata
&\nodata
&$-0.9\pm1$
&GB6 J0745+1011
&
\\
119
&$210.55$
&\phs$54.30$
&$1.7\pm0.1$
&$1.5\pm0.2$
&\nodata
&\nodata
&\nodata
&$-0.3\pm1$
&GB6 J1014+2301
&
\\
120
&$211.25$
&\phs$19.14$
&\nodata
&$1.7\pm0.3$
&$1.2\pm0.2$
&\nodata
&\nodata
&$-1.5\pm3$
&GB6 J0757+0956
&
\\
121
&$212.96$
&\phs$30.09$
&$1.6\pm0.2$
&$2.1\pm0.3$
&\nodata
&\nodata
&\nodata
&$0.6\pm1$
&GB6 J0840+1312
&
\\
122
&$215.23$
&$-24.23$
&$1.4\pm0.1$
&$1.9\pm0.2$
&$1.4\pm0.2$
&\nodata
&\nodata
&$0.3\pm0.6$
&PMN J0527-1241
&
\\
123
&$216.70$
&$-54.26$
&$1.0\pm0.09$
&$1.5\pm0.2$
&$1.2\pm0.2$
&$1.0\pm0.3$
&\nodata
&$0.3\pm0.6$
&PMN J0329-2357
&*
\\
124
&$217.01$
&\phs$11.36$
&$1.8\pm0.1$
&$1.8\pm0.2$
&$1.9\pm0.3$
&$2.3\pm0.5$
&$1.5\pm0.7$
&$0.1\pm0.4$
&GB6 J0739+0136
&
\\
125
&$221.13$
&\phs$22.37$
&$1.4\pm0.1$
&$1.7\pm0.2$
&$1.1\pm0.3$
&$2.0\pm0.5$
&\nodata
&$0.2\pm0.5$
&GB6 J0825+0309
&
\\
126
&$222.59$
&$-16.18$
&$3.9\pm0.1$
&$3.3\pm0.2$
&$3.9\pm0.3$
&$2.0\pm0.4$
&\nodata
&$-0.3\pm0.2$
&PMN J0609-1542
&
\\
127
&$223.45$
&$-30.67$
&$1.1\pm0.09$
&$1.0\pm0.1$
&$1.4\pm0.4$
&\nodata
&\nodata
&$0.1\pm0.8$
&PMN J0513-2159
&
\\
128
&$223.67$
&$-34.89$
&$3.3\pm0.09$
&$3.8\pm0.2$
&$3.8\pm0.2$
&$3.5\pm0.5$
&$2.7\pm0.6$
&$0.2\pm0.2$
&PMN J0457-2324
&
\\
129
&$224.51$
&$-50.79$
&$1.4\pm0.08$
&$1.2\pm0.1$
&$1.0\pm0.1$
&\nodata
&\nodata
&$-0.6\pm0.5$
&PMN J0348-2749
&
\\
130
&$228.66$
&$-13.61$
&$1.3\pm0.1$
&$1.3\pm0.2$
&$1.3\pm0.2$
&\nodata
&\nodata
&$0.1\pm0.6$
&PMN J0629-1959
&
\\
131
&$229.01$
&$-36.96$
&$1.7\pm0.1$
&$1.9\pm0.2$
&$7.6\pm3$
&\nodata
&\nodata
&$0.6\pm0.7$
&PMN J0453-2807
&
\\
132
&$229.00$
&\phs$30.93$
&$2.2\pm0.1$
&$2.1\pm0.2$
&$2.2\pm0.3$
&$0.8\pm0.4$
&\nodata
&$-0.2\pm0.5$
&GB6 J0909+0121
&
\\
133
&$229.04$
&\phs$13.16$
&$1.3\pm0.1$
&$1.4\pm0.2$
&$1.4\pm0.2$
&$2.2\pm0.4$
&$1.2\pm0.6$
&$0.3\pm0.4$
&PMN J0808-0751
&
\\
134
&$229.84$
&$-12.38$
&$1.2\pm0.1$
&$0.3\pm0.1$
&$1.2\pm0.2$
&\nodata
&\nodata
&$-0.2\pm0.6$
&PMN J0636-2041
&*
\\
135
&$231.22$
&$-13.70$
&$0.9\pm0.1$
&$0.9\pm0.1$
&$1.5\pm0.2$
&\nodata
&$1.7\pm0.5$
&$0.6\pm0.5$
&PMN J0633-2223
&
\\
136
&$237.74$
&$-48.48$
&$2.7\pm0.1$
&$2.9\pm0.2$
&$2.5\pm0.2$
&$2.6\pm0.3$
&$2.1\pm0.5$
&$-0.1\pm0.2$
&PMN J0403-3605
&
\\
137
&$239.76$
&$-69.07$
&$0.9\pm0.07$
&$1.0\pm0.1$
&\nodata
&$0.8\pm0.3$
&$1.1\pm0.4$
&$0.1\pm0.5$
&PMN J0222-3441
&
\\
138
&$240.14$
&$-56.75$
&$11.7\pm0.09$
&$7.1\pm0.1$
&$4.3\pm0.1$
&$1.9\pm0.3$
&$1.4\pm0.5$
&$-1.5\pm0.09$
&1Jy 0320-37
&*
\\
139
&$240.61$
&$-32.72$
&$3.9\pm0.1$
&$3.4\pm0.2$
&$3.2\pm0.2$
&$2.8\pm0.4$
&$2.3\pm0.6$
&$-0.4\pm0.2$
&PMN J0522-3628
&
\\
140
&$240.71$
&$-44.43$
&$1.3\pm0.1$
&\nodata
&$1.1\pm0.2$
&$1.1\pm0.4$
&\nodata
&$-0.2\pm0.6$
&PMN J0424-3756
&
\\
141
&$241.23$
&$-47.90$
&$1.0\pm0.1$
&$1.5\pm0.2$
&$1.3\pm0.2$
&$1.0\pm0.3$
&\nodata
&$0.3\pm0.6$
&PMN J0406-3826
&
\\
142
&$241.72$
&\phs$51.54$
&$1.8\pm0.2$
&$1.8\pm0.3$
&$1.4\pm0.3$
&$1.9\pm0.5$
&\nodata
&$-0.1\pm0.5$
&GB6 J1038+0512
&
\\
143
&$242.86$
&\phs$25.11$
&$2.0\pm0.1$
&\nodata
&$0.8\pm0.3$
&\nodata
&$1.4\pm0.7$
&$-0.5\pm0.7$
&PMN J0918-1205
&
\\
144
&$243.56$
&\phs$12.26$
&$3.3\pm0.1$
&$2.9\pm0.2$
&$3.0\pm0.2$
&$2.1\pm0.3$
&\nodata
&$-0.3\pm0.2$
&PMN J0836-2017
&
\\
145
&$244.36$
&\phs$6.05$
&$1.1\pm0.09$
&$1.4\pm0.2$
&$1.7\pm0.2$
&$2.0\pm0.4$
&\nodata
&$0.6\pm0.4$
&PMN J0816-2421
&
\\
146
&$244.72$
&$-54.06$
&$1.7\pm0.1$
&$1.7\pm0.2$
&$1.7\pm0.2$
&$2.0\pm0.4$
&$2.2\pm0.8$
&$0.1\pm0.3$
&PMN J0334-4008
&
\\
147
&$248.42$
&$-41.55$
&$3.4\pm0.1$
&$3.7\pm0.2$
&$3.4\pm0.2$
&$2.5\pm0.3$
&$2.5\pm0.7$
&$-0.1\pm0.2$
&PMN J0440-4332
&
\\
148
&$250.09$
&$-31.08$
&$7.9\pm0.1$
&$8.2\pm0.2$
&$8.8\pm0.2$
&$7.9\pm0.4$
&$6.7\pm0.6$
&$0.0\pm0.07$
&PMN J0538-4405
&
\\
149
&$251.50$
&\phs$52.77$
&$5.9\pm0.1$
&$5.7\pm0.2$
&$6.5\pm0.3$
&$5.2\pm0.4$
&$4.6\pm2$
&$-0.0\pm0.1$
&GB6 J1058+0133
&
\\
150
&$251.59$
&$-34.64$
&$6.8\pm0.1$
&$5.3\pm0.2$
&$3.9\pm0.2$
&$2.9\pm0.3$
&\nodata
&$-0.8\pm0.1$
&PMN J0519-4546
&*
\\
151
&$251.99$
&$-38.80$
&$3.0\pm0.1$
&$4.0\pm0.2$
&$3.7\pm0.2$
&$3.4\pm0.4$
&$3.8\pm0.9$
&$0.3\pm0.2$
&PMN J0455-4616
&
\\
152
&$262.01$
&$-31.87$
&$1.3\pm0.1$
&$1.4\pm0.2$
&$1.8\pm0.2$
&$1.6\pm0.3$
&\nodata
&$0.3\pm0.4$
&PMN J0540-5418
&
\\
153
&$265.89$
&$-30.67$
&$1.3\pm0.09$
&$3.1\pm1$
&$1.2\pm0.2$
&\nodata
&\nodata
&$-0.0\pm0.6$
&PMN J0550-5732
&
\\
154
&$270.52$
&$-36.04$
&$2.6\pm0.09$
&$2.4\pm0.1$
&$2.0\pm0.2$
&$1.5\pm0.3$
&\nodata
&$-0.4\pm0.3$
&PMN J0506-6109
&*
\\
155
&$272.49$
&$-54.59$
&$2.7\pm0.1$
&$3.1\pm0.1$
&$3.5\pm0.2$
&$2.9\pm0.3$
&$2.2\pm0.5$
&$0.2\pm0.2$
&PMN J0253-5441
&
\\
156
&$273.59$
&$-31.25$
&$0.8\pm0.06$
&$0.6\pm0.09$
&$0.6\pm0.1$
&\nodata
&\nodata
&$-0.6\pm0.8$
&PMN J0546-6415
&
\\
157
&$275.35$
&\phs$43.60$
&$1.9\pm0.1$
&$1.7\pm0.2$
&$1.9\pm0.2$
&$1.4\pm0.3$
&\nodata
&$-0.2\pm0.4$
&PMN J1130-1449
&
\\
158
&$276.09$
&$-61.76$
&$3.0\pm0.1$
&$2.9\pm0.2$
&$3.0\pm0.2$
&$3.4\pm0.4$
&$1.8\pm0.6$
&$0.0\pm0.2$
&PMN J0210-5101
&
\\
159
&$276.76$
&\phs$39.62$
&\nodata
&$0.7\pm0.2$
&$1.2\pm0.2$
&$1.3\pm0.5$
&\nodata
&$1.0\pm2$
&PMN J1127-1857
&
\\
160
&$278.14$
&$-48.90$
&$1.3\pm0.1$
&\nodata
&\nodata
&$1.6\pm0.5$
&\nodata
&$0.2\pm0.8$
&PMN J0309-6058
&
\\
161
&$279.51$
&$-20.16$
&$1.6\pm0.1$
&$1.0\pm0.1$
&$1.0\pm0.2$
&\nodata
&\nodata
&$-1.1\pm0.6$
&PMN J0743-6726
&*
\\
162
&$280.29$
&$-48.71$
&$1.5\pm0.1$
&$1.6\pm0.2$
&$1.5\pm0.2$
&$1.7\pm0.3$
&\nodata
&$0.1\pm0.4$
&PMN J0303-6211
&
\\
163
&$281.44$
&\phs$9.75$
&$1.2\pm0.1$
&$1.1\pm0.2$
&$0.5\pm0.1$
&\nodata
&\nodata
&$-0.9\pm1$
&PMN J1041-4740
&
\\
164
&$281.86$
&\phs$67.35$
&$2.6\pm0.1$
&$2.1\pm0.2$
&$1.9\pm0.2$
&$2.0\pm0.6$
&\nodata
&$-0.5\pm0.4$
&GB6 J1219+0549A
&*
\\
165
&$283.81$
&\phs$74.49$
&$19.7\pm0.1$
&$15.8\pm0.2$
&$13.1\pm0.2$
&$9.2\pm0.4$
&$5.3\pm0.6$
&$-0.7\pm0.06$
&GB6 J1230+1223
&
\\
166
&$284.19$
&\phs$14.25$
&$2.0\pm0.09$
&$1.7\pm0.1$
&$1.4\pm0.1$
&$1.4\pm0.4$
&$1.7\pm0.7$
&$-0.4\pm0.3$
&PMN J1107-4449
&
\\
167
&$286.39$
&$-27.15$
&$4.6\pm0.09$
&$4.4\pm0.1$
&$3.9\pm0.1$
&$3.6\pm0.3$
&\nodata
&$-0.3\pm0.1$
&PMN J0635-7516
&
\\
168
&$288.31$
&$-63.89$
&\nodata
&$1.3\pm0.2$
&$0.5\pm0.1$
&\nodata
&\nodata
&$-4.8\pm4$
&PMN J0133-5159
&
\\
169
&$289.25$
&\phs$22.95$
&$2.9\pm0.1$
&$3.3\pm0.2$
&$2.9\pm0.2$
&$2.8\pm0.4$
&\nodata
&$0.0\pm0.2$
&PMN J1147-3812
&
\\
170
&$289.95$
&\phs$64.36$
&$20.0\pm0.1$
&$18.3\pm0.2$
&$17.5\pm0.3$
&$14.5\pm0.4$
&$9.0\pm0.8$
&$-0.3\pm0.04$
&GB6 J1229+0202
&
\\
171
&$290.72$
&$-76.16$
&$2.3\pm0.09$
&$2.4\pm0.2$
&$1.9\pm0.2$
&$1.6\pm0.4$
&\nodata
&$-0.2\pm0.3$
&PMN J0106-4034
&
\\
172
&$290.54$
&\phs$37.82$
&$1.5\pm0.1$
&$1.1\pm0.2$
&\nodata
&\nodata
&\nodata
&$-0.9\pm1$
&PMN J1209-2406
&
\\
173
&$291.04$
&\phs$44.52$
&$1.7\pm0.1$
&$1.6\pm0.3$
&$1.6\pm0.3$
&$1.3\pm0.4$
&\nodata
&$-0.2\pm0.5$
&PMN J1215-1731
&
\\
174
&$293.30$
&$-37.61$
&$1.0\pm0.09$
&$1.3\pm0.1$
&$1.0\pm0.3$
&$1.3\pm0.5$
&\nodata
&$0.4\pm0.6$
&PMN J0311-7651
&
\\
175
&$293.85$
&$-31.40$
&$2.0\pm0.1$
&$2.2\pm0.2$
&$1.8\pm0.1$
&$2.5\pm0.4$
&$1.6\pm0.7$
&$0.1\pm0.3$
&PMN J0450-8100
&
\\
176
&$298.06$
&$-18.27$
&$2.0\pm0.1$
&$2.2\pm0.1$
&$2.4\pm0.2$
&$2.4\pm0.3$
&\nodata
&$0.2\pm0.3$
&PMN J1058-8003
&
\\
177
&$301.63$
&\phs$37.12$
&$1.4\pm0.2$
&$1.9\pm0.4$
&\nodata
&$1.0\pm0.4$
&\nodata
&$-0.0\pm0.8$
&PMN J1246-2547
&
\\
178
&$302.00$
&$-20.33$
&$1.1\pm0.09$
&$1.6\pm0.8$
&$1.3\pm0.1$
&\nodata
&\nodata
&$0.3\pm0.5$
&PMN J1224-8312
&
\\
179
&$303.32$
&$-59.45$
&$1.3\pm0.1$
&$1.2\pm0.2$
&$0.7\pm0.1$
&$1.5\pm0.4$
&\nodata
&$-0.2\pm0.5$
&PMN J0050-5738
&
\\
180
&$304.58$
&\phs$30.87$
&$1.7\pm0.1$
&$1.3\pm0.2$
&$1.7\pm0.3$
&\nodata
&\nodata
&$-0.3\pm0.6$
&PMN J1257-3154
&
\\
181
&$305.11$
&\phs$57.06$
&$23.5\pm0.1$
&$24.9\pm0.2$
&$25.7\pm0.3$
&$24.5\pm0.5$
&$19.0\pm0.9$
&$0.1\pm0.03$
&PMN J1256-0547
&
\\
182
&$308.79$
&\phs$29.02$
&$1.4\pm0.1$
&$1.6\pm0.2$
&$1.7\pm0.2$
&$1.6\pm0.4$
&\nodata
&$0.2\pm0.4$
&PMN J1316-3339
&
\\
183
&$313.45$
&$-18.85$
&$2.4\pm0.1$
&$2.0\pm0.1$
&$1.8\pm0.2$
&$2.0\pm0.4$
&\nodata
&$-0.4\pm0.3$
&PMN J1617-7717
&
\\
184
&$313.47$
&$-35.13$
&$1.6\pm0.09$
&$1.3\pm0.1$
&$1.0\pm0.2$
&\nodata
&\nodata
&$-0.6\pm0.6$
&PMN J2146-7755
&
\\
185
&$313.54$
&\phs$27.99$
&$1.7\pm0.1$
&$1.5\pm0.2$
&$1.1\pm0.2$
&$1.7\pm0.4$
&\nodata
&$-0.3\pm0.5$
&PMN J1336-3358
&
\\
186
&$313.68$
&$-23.40$
&$1.0\pm0.1$
&\nodata
&$1.0\pm0.1$
&\nodata
&\nodata
&$0.0\pm0.7$
&PMN J1733-7935
&
\\
187
&$314.12$
&$-55.14$
&$1.8\pm0.1$
&\nodata
&$1.3\pm0.1$
&\nodata
&\nodata
&$-0.5\pm0.5$
&PMN J2358-6054
&
\\
188
&$320.02$
&\phs$48.37$
&$6.3\pm0.1$
&$6.7\pm0.2$
&$7.3\pm0.3$
&$6.6\pm0.4$
&$3.7\pm0.7$
&$0.1\pm0.1$
&PMN J1337-1257
&
\\
189
&$320.14$
&$-62.06$
&\nodata
&$1.3\pm0.2$
&\nodata
&$1.7\pm0.6$
&\nodata
&$0.4\pm1$
&PMN J2357-5311
&
\\
190
&$321.31$
&$-40.64$
&$3.6\pm0.1$
&$2.9\pm0.2$
&$2.6\pm0.2$
&$2.2\pm0.4$
&\nodata
&$-0.6\pm0.2$
&PMN J2157-6941
&
\\
191
&$321.44$
&\phs$17.27$
&$2.6\pm0.1$
&$3.0\pm0.2$
&$2.6\pm0.3$
&\nodata
&\nodata
&$0.1\pm0.4$
&PMN J1427-4206
&
\\
192
&$323.74$
&$-24.44$
&$2.0\pm0.09$
&$1.9\pm0.1$
&$1.5\pm0.1$
&$1.7\pm0.3$
&\nodata
&$-0.3\pm0.3$
&PMN J1837-7108
&
\\
193
&$325.13$
&\phs$25.60$
&$0.8\pm0.1$
&$1.7\pm0.2$
&$1.9\pm0.2$
&$2.4\pm0.6$
&\nodata
&$1.3\pm0.6$
&PMN J1427-3306
&*
\\
194
&$326.21$
&$-34.56$
&$1.0\pm0.2$
&$1.3\pm0.2$
&$0.9\pm0.2$
&\nodata
&$2.0\pm0.7$
&$0.4\pm0.6$
&PMN J2035-6846
&
\\
195
&$326.81$
&$-60.81$
&$0.9\pm0.08$
&$0.8\pm0.2$
&\nodata
&\nodata
&\nodata
&$-0.6\pm2$
&PMN J2336-5236
&
\\
196
&$327.01$
&$-15.87$
&$2.1\pm0.1$
&$2.0\pm0.2$
&$1.6\pm0.2$
&\nodata
&\nodata
&$-0.3\pm0.4$
&PMN J1723-6500
&
\\
197
&$327.19$
&\phs$49.19$
&$1.5\pm0.1$
&$1.2\pm0.2$
&$2.0\pm0.4$
&$2.6\pm0.6$
&\nodata
&$0.4\pm0.5$
&PMN J1354-1041
&
\\
198
&$328.10$
&$-12.44$
&$1.9\pm0.1$
&$2.0\pm0.2$
&$2.3\pm0.2$
&$2.0\pm0.3$
&\nodata
&$0.1\pm0.3$
&PMN J1703-6212
&
\\
199
&$328.83$
&$-19.57$
&$1.2\pm0.1$
&$1.4\pm0.2$
&$1.5\pm0.2$
&$1.2\pm0.3$
&\nodata
&$0.2\pm0.5$
&PMN J1803-6507
&
\\
200
&$330.86$
&$-20.79$
&$1.7\pm0.1$
&$1.2\pm0.2$
&$1.1\pm0.2$
&$1.4\pm0.5$
&\nodata
&$-0.7\pm0.5$
&PMN J1819-6345
&
\\
201
&$331.72$
&$-52.11$
&$1.0\pm0.07$
&\nodata
&\nodata
&\nodata
&$1.4\pm0.6$
&$0.2\pm0.7$
&PMN J2239-5701
&
\\
202
&$332.58$
&$-74.91$
&$1.5\pm0.1$
&$1.2\pm0.2$
&$0.9\pm0.2$
&$1.0\pm0.3$
&\nodata
&$-0.6\pm0.5$
&PMN J0013-3954
&
\\
203
&$333.90$
&\phs$50.35$
&$1.5\pm0.2$
&$0.8\pm0.2$
&$1.5\pm0.3$
&\nodata
&\nodata
&$-0.3\pm0.9$
&1Jy 1406-076
&
\\
204
&$334.81$
&$-60.55$
&$1.2\pm0.08$
&$1.2\pm0.1$
&$1.2\pm0.2$
&\nodata
&\nodata
&$0.0\pm0.5$
&PMN J2315-5018
&
\\
205
&$340.68$
&\phs$27.58$
&$1.9\pm0.1$
&$2.0\pm0.2$
&$2.0\pm0.3$
&$1.9\pm0.6$
&\nodata
&$0.1\pm0.5$
&PMN J1517-2422
&
\\
206
&$344.52$
&$-56.10$
&$2.0\pm0.1$
&$1.8\pm0.2$
&$2.0\pm0.2$
&$2.6\pm0.4$
&\nodata
&$0.1\pm0.3$
&PMN J2235-4835
&
\\
207
&$351.26$
&\phs$40.16$
&$1.9\pm0.1$
&$1.9\pm0.2$
&$1.8\pm0.3$
&\nodata
&$1.7\pm0.7$
&$-0.1\pm0.5$
&1Jy 1510-08
&
\\
208
&$352.48$
&$-40.31$
&$1.3\pm0.1$
&\nodata
&\nodata
&$1.7\pm0.5$
&\nodata
&$0.2\pm0.7$
&PMN J2056-4714
&
\\
\enddata
\tablecomments{
* indicates that the source has two possible identifications.  
          Source 138 (Fornax A) is extended; for this source only 
          we relaxed our requirement that the width of the fitted Gaussian 
          profile be within $5\sigma$ of the beam width.
}
\end{deluxetable}

\clearpage

\section{CONCLUSIONS \label{conclusions}}

1) We have defined and constructed a series of ``standard'' masks to use for CMB data 
analyses.  The masks vary in degree of the severity of the sky cuts.  These masks have 
been used in the companion \iMAP\ papers that analyze the CMB anisotropy.

2) We present a CMB map formed from a weighted combination of the five \iMAP\ bands designed to 
reduce foreground contamination.  The method is effective, but care must be used in 
the use of this map given its complicated noise correlations.

3) We examine the fitting of foreground spatial template maps to the \iMAP\ maps to remove 
foregrounds for CMB analyses.  This method can 
work well as long a mask is applied to exclude the strongest regions of Galactic emission 
and as long as the fits are restricted in frequency.  These restrictions are 
needed because the foreground 
tracers are not high fidelity tracers of the \iMAP\ forgrounds.  A 
frequency-restricted fit with a mask is used for the current ``first year'' \iMAP\ results.

4) The relatively flat synchrotron spectral index ($\beta\approx -2.5$) in the Galactic plane 
between 408 MHz and 23 GHz provides evidence for diffusion and convection of 
cosmic rays within the Galactic plane.  
The significant steepening of the synchrotron spectral index at high Galactic latitudes 
suggests that diffusion is the dominant process for cosmic ray electrons leaving the plane, 
and that these cosmic ray electrons are trapped in the halo for sufficient time to lose 
a significant amount of their energy.

5) We have made maps and derived spectral indices of the individual physical emission 
components using a Maximum Entropy Method. 

6) The steep spectral index ($\beta\approx -3$) high Galactic latitude features, seen prominantly 
in low frequency radio maps, become comparatively weaker than the flatter ($\beta\approx -2.5$) 
synchrotron emission in the plane at the higher \iMAP\ frequencies.  The consequence of this 
is that the synchrotron sky map seen by \iMAP\ is a much closer tracer of regions of 
recent star-formation than low radio frequency maps.  
This, in turn, implies that the spatial distribution of synchrotron 
emission at the \iMAP\ frequencies is strongly correlated with that of other 
star-formation tracers, such as thermal dust emission.  
The synchrotron spectral index is steeper
in the \iMAP\ bands than at radio frequencies.

7) A synchrotron map of the sky is frequency dependent.    
It is inappropriate to assume that 
a synchrotron map measured at low frequencies (e.g., 408 MHz) should be a high-fidelity 
tracer of synchrotron emission at much higher frequencies.

8) The free-free map from \iMAP\ is generally consistent with H$\alpha$ measurements 
that also trace the ionized gas.

9) The spectral index of the thermal dust emission is steep, $\beta_d\approx2.2$ in the 
\iMAP\ bands, possibly implying emission from cold silicate grains.

10) There is no frequency at which spinning or magnetic dust emission dominates in the \iMAP\ 
data.  Spinning dust emission is $<5$ \%
of the Ka-band antenna temperature.

11) The \iMAP\ bands were chosen to be in a region of the spectrum where the ratio of the 
CMB to foreground contamination was at a maximum.  Due to the steep synchrotron and dust 
spectra, the choice of \iMAP\ frequencies was even more advantageous than expected.

12) The total flux of our Galaxy in the five \iMAP\ bands is 45, 38, 35, 32, and  
48 kJy, from K-band to W-band.

13) The Sunyaev-Zeldovich effect is a negligible ``contaminant'' for the current first 
year results, as expected.

14) A catalog of 208 point sources detected by \iMAP\ is presented.  Of these, five are likely
to be spurious.  The mean spectral index of the point sources  
is $\alpha \sim 0$ ($\beta\sim -2$).  Derived source counts suggest a contribution to the anisotropy 
power of $(15.0\pm 1.4)\times 10^{-3}$ $\mu$K$^2$sr at Q-band and negligible levels at V-band and 
W-band.

\acknowledgements

The \iMAP\ mission is made possible by the support of the Office of Space 
Sciences at NASA Headquarters and by the hard and capable work of scores of 
scientists, engineers, technicians, machinists, data analysts, budget analysts, 
managers, administrative staff, and reviewers.  The Wisconsin H-Alpha Mapper 
(WHAM), the Virginia Tech Spectral-Line Survey (VTSS), and the 
Southern H-Alpha Sky Survey Atlas (SHASSA), 
are funded by the National Science Foundation.  We are grateful to D. Finkbeiner
for making his combined H$\alpha$ map available to us in advance of publication.

\end{document}